\documentclass[10pt]{article}
\usepackage{amssymb}
\usepackage{epsfig}
\input

\begin{document}

\title{Estimation of Stress–-Strength model in the Generalized Linear Failure Rate Distribution}
\author{{\textbf{ Fatemeh Shahsanaei$^{\dag}$ and Alireza Daneshkhah$^{\ddag}$} }\\
$\dag$Department of Statistics, Faculty of Mathematical Sciences and Computer\\
Shahid Chamran University, Ahvaz 6135714463, Iran\\
$\ddag$ Cranfield Water Science Institute, School of Applied Sciences\\
Cranfield University, Cranfield, MK43 0AL, UK\\
${\ddag}$fatemehshahsanaie@yahoo.com; $\ddag$ a.daneshkhah@cranfield.ac.uk}
\maketitle

\begin{abstract}
In this paper, we study the estimation of $R=P [Y < X ]$, also so-called the stress-strength model, when both $X$ and $Y$ are
two independent random variables with the generalized linear failure rate distributions, under different assumptions about their parameters. We address the maximum likelihood estimator (MLE) of $R$ and the associated asymptotic confidence
interval. In addition, we compute the MLE and the corresponding Bootstrap confidence interval when the sample sizes are small. The Bayes estimates of $R$ and the associated credible intervals are also investigated. An extensive computer simulation is implemented to compare the performances of the proposed estimators. Eventually, we briefly study the estimation of this model when the data obtained from both distributions are progressively type-II censored. We present the MLE and the corresponding confidence interval under three different progressive censoring schemes. We also analysis a set of real data for illustrative purpose.
\end{abstract}


\section{Introduction}
The topic of inference on $R=P(Y<X)$ - usually referred to as the stress-strength model - has
obtained wide attention in the literature, including quality control, engineering statistics, reliability, medicine, psychology, biostatistics, stochastic precedence, and probabilistic mechanical design (see Kotz et al., 2003, for a comprehensive review).  For instance, in a clinical study, $Y$ and $X$ can be assumed as the outcomes of a treatment and a control group, respectively, then the following quantity $R = P(Y< X )$ can be considered as the effectiveness of the treatment (Kotz et al., 2003). In this case, $(1-R)$ measures the effectiveness of the treatment. Alternatively,
for diagnostic tests used to distinguish between diseased and non-diseased patients, the area under the receiver operating characteristics (ROC) curve, based on the sensitivity and the complement to specificity at different cut-off points of the range of possible test
values, is equal to $R$ (see Ventura and Racugno, 2011).
\par
Another important use of $R = P(Y < X )$ is in reliability contexts, in particular in mechanical reliability of a system, where $Y$ is the strength of a component which is subject to stress $X$, then $R$ is a measure of system performance, and $(1-R)$ measures the chance that
the system fails. In this situation, the system will fail, if at any time the applied stress is greater than its strength. Kotz et al. (2003) also present the theoretical and practical results on the theory and applications of the stress–strength relationships in industrial and economic systems.
\par
In reliability context and life science, inferences about $R$ where $X$ and $Y$ are independently distributed are still subject of interest. In this context, the stress-strength model describes the life of a component which has a random strength $X$ and is subjected to random stress $Y$. The component fails at the instant that the stress applied to it exceeds the strength and the component will function satisfactorily whenever $Y<X$ . Thus $R = P(Y<X)$ is a measure of component reliability.
\par
Estimation of $R = P(Y<X)$, when $X$ and $Y$ are random variables following the specified distributions has been extensively discussed in the literature
in both parametric and non-parametric framework. This quantity can be obviously seen as a function of the parameters of the distribution of the random vector $(X, Y)$ and could be calculated in the closed form for a limited number of cases (Kotz et al., 2003; Nadarajah, 2005; Cordeiro et al, 2011). For instance, the estimation of $R$  when $X$ and $Y$ are independent and normally distributed has been considered by several authors including Downtown (1973), Owen et al. (1977) and Greco and Ventura (2011).
\par
Recently, Rezaei et al. (2010) reported a list of papers related to the estimation problem of $R$ when $X$ and $Y$ are independent and follow a class of life-time distributions including Exponential, bivariate Exponential, generalized exponential, Gamma distributions, Burr type $X$ model, Weibull distribution, and among others. 
\par
In this paper, the main objective of this paper is to focus on the inference of $R = P[Y < X]$, where $X$ and $Y$ follow the \textit{Generalized Linear Failure Rate} distributions and are independent of each other. This distribution is originally introduced by Sarhan and Kundu (2007). Similar to the other studies, we first obtain the MLE of $R$ and its corresponding asymptotic distribution. We then construct an asymptotic confidence interval based on the asymptotic distribution. In addition, we present a Bootstrap confidence interval for $R$ when the sample sizes are small. We also derive the Bayes estimates of $R$ associated with the informative and non-informative prior distributions, and the associated credible intervals are also calculated.
\par
Furthermore, we briefly investigate the statistical inference of the stress-strength parameter $R =P(X < Y )$ when the observe sample from $X$ and $Y$ are progressively type-II censored. We only calculate the MLEs and associated confidence intervals for three progressive censoring schemes and further studies about $R$ under these censoring schemes will be reported later.
\par
The rest of the paper is organized as follows. We briefly introduce the {Generalized Linear Failure Rate} (GLFR) distribution and study its relevant properties to this study in Section \ref{GLFR}. We devote Section 3 to study the estimation of $R$ when the scale parameters of both distributions are common and known. In this section, we derive the ML estimator, Bayes estimators of the stress-strength model, their corresponding confidence or credible intervals and other quantities of interests. In Section 4, we carry out similar inference, made in the previous section, about $R$ when the common scale
parameters are unknown is discussed in Section 4. We consider inference about $R$ for the general case when the parameters of both distributions are not known and common in Section 5. We derive maximum likelihood estimators of $R$ and its corresponding confidence intervals under different progressive censoring schemes in Sections 6. Simulation results and data analysis are presented in Sections 7 and 8, respectively.
\bigskip

\section{Generalized Linear Failure Rate Distribution}\label{GLFR}

It is well known that the exponential, generalized exponential or Rayleigh distribution are among the most commonly used distributions for
analyzing lifetime data. These distributions have several desirable properties and nice
physical interpretations. They can be used quite effectively in modelling strength and general lifetime data. Kundu and Raqab (2005) used different methods to estimate the parameters of the generalized Rayleigh on the observed data. In analyzing lifetime data, the exponential, Rayleigh, linear failure rate or generalized exponential distributions are normally used. It is apparent that the exponential distribution can be only used for the constant hazard function whereas Rayleigh, linear failure rate and generalized exponential distributions can be used for the monotone (increasing in case of Rayleigh or linear failure rate and increasing/ decreasing in case of generalized exponential distribution) hazard functions. In addition, in many practical applications, it is required to apply the non-monotonic function such as bathtub shaped hazard function (Lai et al. 2001). In this paper we use a newly developed distribution by Sarhan and Kundu (2007) which generalizes the well known exponential distribution, linear failure rate distribution, generalized exponential distribution, and generalized Rayleigh distribution (also known as Burr Type $X$ distribution). They called it \textit{generalized linear failure rate} distribution with three parameters $(a, b,\alpha)$ and denoted by GLFRD$(a, b,\alpha)$. The probability density function (pdf) of $GLFRD(a,b,\beta)$ is given by
\[
f_{X}(a,b,\alpha)(x)=\alpha(a+bx)e^{-(ax+\frac{b}{2}x^{2})}(1-e^{-(ax+\frac{b}{2}x^{2})})^{\alpha-1}
 \hspace{3mm} ; a,b,\alpha > 0~~~x>0
\]
The corresponding cumulative distribution function is as follows
\begin{equation}
F_{X}(x)=(1-e^{-(ax+\frac{b}{2}x^{2})})^{\alpha}\label{ge-Lin-fail-CDF}
\end{equation}
where $a$ and $b$ are the scale parameters and $\alpha$ is the shape parameter.
\par
This distribution
has increasing, decreasing or bathtub shaped hazard rate functions and it also generalizes many well known distributions
including the traditional linear failure rate distributions, such as, generalized exponential $(GED(a,\alpha))$ and generalized Rayleigh  $(GRD(b,\alpha))$ by putting $b=0$ and $a =0$, respectively.
\par
\begin{figure}[h]
\begin{center}
\scalebox{0.6}{\includegraphics{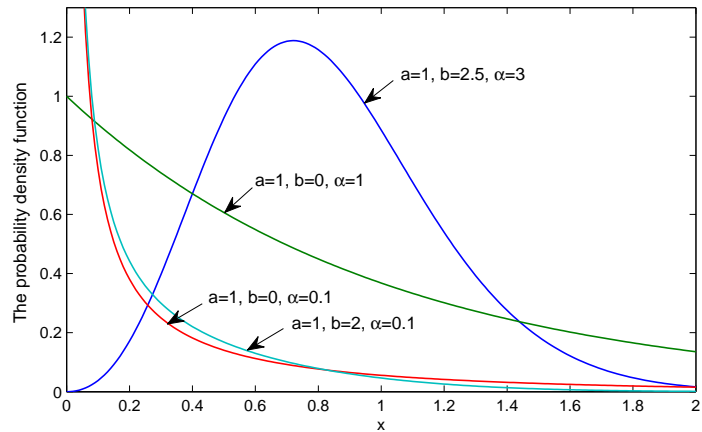}}
\end{center}
\caption{Different shapes of pdf of the GLFR distribution, including unimodal pdf}\label{pdf}
\end{figure}
\begin{figure}[h]
\begin{center}
\scalebox{0.6}{\includegraphics{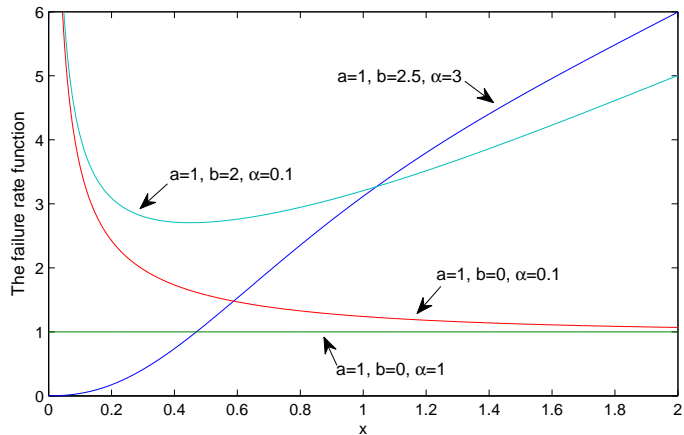}}
\end{center}
\caption{Different shapes of hazard rate function of the GLFR distribution}\label{hazard}
\end{figure}

This distribution is verified to have a decreasing or unimodal pdf. Figure {\ref{pdf}} shows some patterns of the pdf of $GLFRD(a,b,\alpha)$, which may have a single mode or no mode at all.
In addition, when $\alpha > 1$, the hazard rate of this distribution is increasing, if $\alpha < 1$, the associated hazard rate is either decreasing
if $b = 0$ or inverted bathtub if $b> 0$, and finally when $\alpha = 1$, the hazard rate is either increasing if $b > 0$ or constant if $b = 0$. These patterns are shown in Figure \ref{hazard} for differen values of the parameters.
\par
Sarhan et al (2008) studied the statistical properties of this distribution and provided some nice physical interpretations. The maximum likelihood estimates (MLEs) of the corresponding parameters appeared to not have the explicit forms, and they can be obtained only by solving two non-linear equations.
\par

\section{Estimation of $R$ with known scale parameters}
In this section, the main aim is the estimation of $R = P[Y<X]$, where independent random variables $X$ and $Y$ follow the \textit{Generalized Linear Failure Rate} distributions with the known common scale parameters, that is, $X\sim GLFRD(a,b,\alpha)$ and $Y\sim GLFRD(a,b,\beta)$. We wish to derive the MLE of $R$, its associated confidence intervals, Bayes estimates of $R$, the corresponding credible interval and study their properties. The stress-strength parameter, $R$ is defined as
\[
R = P [Y < X ] = \int_{0}^{\infty} P( Y < X |X = x )f_{X}(x)dx
\]
\begin{equation}
=\int_{0}^{\infty}\alpha(a+bx)e^{-(ax+\frac{b}{2}x^{2})}(1-e^{-(ax+\frac{b}{2}x^{2})})^{\alpha-1}(1-e^{-(ax+\frac{b}{2}x^{2})})^{\beta}dx\\
=\frac{\alpha}{\alpha+\beta}\label{R}
\end{equation}
\subsection{MLE of $R$}
In this section, we consider the estimation of $R$ when $(a, b)$ are
known, and without loss of generality, we assume that
$(a,b)=(1,2)$. Let $X_{1},X_{2},\ldots,X_{n}$ be a random sample from
$GLFR(1,2,\alpha)$ and $Y_{1},Y_{2},\ldots,Y_{m}$
 be a random sample from $GLFR(1,2,\beta)$. To compute the MLE of $R$, the corresponding log-likelihood of the observed sample is given by
\[
\ell(\alpha,\beta)=n\ln\alpha+\sum_{i=1}^{n}\ln(1+2x_{i})+(\alpha-1)\sum_{i=1}^{n}\ln(1-e^{-(x_{i}+x_{i}^{2})})-\sum_{i=1}^{n}(x_{i}+x_{i}^{2})
\]
\begin{equation}
+m\ln\beta+\sum_{j=1}^{m}\ln(1+2y_{j})+(\beta-1)\sum_{j=1}^{m}\ln(1-e^{-(y_{j}+y_{j}^{2})}) -\sum_{j=1}^{m}(y_{j}+y_{j}^{2})\label{loglike}
\end{equation}

The MLEs of ($\alpha$, $\beta$) denoted by ($\hat{\alpha}$, $\hat{\beta}$) can be derived by solving the following equations
\[
\frac{\partial \ell}{\partial
\alpha}=\frac{n}{\alpha}+\sum_{i=1}^{n}\ln(1-e^{-(x_{i}+x_{i}^{2})})
\]
\[
\frac{\partial \ell}{\partial
\beta}=\frac{m}{\beta}+\sum_{j=1}^{m}\ln(1-e^{-(y_{j}+y_{j}^{2})})
\]
Consequently, ($\hat{\alpha}$, $\hat{\beta}$) are given by
\[
\hat{\alpha}=\frac{-n}{\sum_{i=1}^{n}\ln(1-e^{-(x_{i}+x_{i}^{2})})}
\]
\[
\hat{\beta}=\frac{-m}{\sum_{j=1}^{m}\ln(1-e^{-(y_{j}+y_{j}^{2})})}
\]
Duo to the invariant property of maximum likelihood estimators, the MLE of $R$ is obtained by replacing $\alpha$ and $\beta$ by their MLEs in (\ref{R}) as follows
\[
\hat{R}=\frac{\hat{\alpha}}{\hat{\alpha}+\hat{\beta}}
\]
Therefore,
\[
\hat{R}=\frac{n\sum_{j=1}^{m}\ln(1-e^{-(y_{j}+y_{j}^{2})})}{n\sum_{j=1}^{m}\ln(1-e^{-(y_{j}+y_{j}^{2})})+m\sum_{i=1}^{n}\ln(1-e^{-(x_{i}+x_{i}^{2})})}\label{RhatF1}
\]
It is trivial to show that $-\ln(1-e^{-(X_{i}+X_{i}^{2})})$ follows an exponential distribution with mean $\alpha^{-1}$. Therefore, $-2\alpha\sum_{i=1}^{n}\ln(1-e^{-(X_{i}+X_{i}^{2})})\sim{\chi^{2}_{(2n)}}$
and
$-2\beta\sum_{j=1}^{m}\ln(1-e^{-(Y_{j}+Y_{j}^{2})})\sim{\chi^{2}_{(2m)}}$.
So,
$$\hat{R}\sim\frac{1}{1+\frac{\beta}{\alpha}F}$$
or
$$\frac{R}{1-R}\times\frac{1-\hat{R}}{\hat{R}}\sim F,$$
where the random variable F follows a $F_{(2n,2m)}$ distribution
with $2n$ and $2m$ degrees of freedom. So, the probability density
function (pdf) of $\hat{R}$ is as follows:
\[
f_{\hat{R}}(x)=\frac{1}{x^{2}B(n,m)}(\frac{n\alpha}{m\beta})^{n}\times\frac{(\frac{1-x}{x})^{n-1}}{(1+\frac{n\alpha}{m\beta}(\frac{1-x}{x}))^{n+m}},
\]
where $0<x<1$ and $\alpha,\beta>0$. The $100(1 -\gamma)\%$ confidence interval of $R$ can be obtained as
\[
[\frac{1}{1+F_{(1-\frac{\gamma}{2};2m,2n)}\times(\frac{1}{\hat{R}}-1)},\frac{1}{1+F_{(\frac{\gamma}{2};2m,2n)}\times(\frac{1}{\hat{R}}-1)}]
\]
where $F_{(\frac{\gamma}{2};2m,2n)}$ and
$F_{(1-\frac{\gamma}{2};2m,2n)}$ are the lower and upper
$\frac{\gamma}{2}${th} percentile points of a F distribution.

\subsection{Bayes estimation of R}
Let $X\sim GLFR(1,2,\alpha)$ and $Y\sim GLFR(1,2,\beta)$ be independent random variables with cumulative distribution functions $F_{X}(x\mid \alpha)$ and $F_{Y}(y\mid \beta)$ given in (\ref{ge-Lin-fail-CDF}), respectively. By definition, $R$ can be evaluated as a function of the entire parameter $\theta=(\alpha,\beta)$, by the following relation
\[
R=R(\theta)=P(X<Y)=\int{F_{X}(t\mid \alpha)f_{Y}(t\mid \beta)dt}
\]
where $\textbf{x}=(x_{1},\ldots,x_{n})$ is a random sample of size $n$ from $X$ and $\textbf{y}=(y_{1},\ldots,y_{m})$ is a random sample of size $m$ from $Y$. Let $\pi(\theta)=\pi(\alpha)\pi(\beta)$ be a prior pdf on $(\alpha, \beta)$. We consider the Gamma distributions as the prior distributions on $\alpha$ and $\beta$, that is, $\alpha\sim Gamma(\gamma_{1},\lambda_{1})$ and
$\beta\sim Gamma(\gamma_{2},\lambda_{2})$, with the following density function, respectively
\begin{equation}
\pi(\alpha)=\frac{\lambda_{1}^{\gamma_{1}}}{\Gamma(\gamma_{1})}\alpha^{\gamma_{1}-1}e^{-\lambda_{1}\alpha},~~~
\pi(\beta)=\frac{\lambda_{2}^{\gamma_{2}}}{\Gamma(\gamma_{2})}\beta^{\gamma_{2}-1}e^{-\lambda_{2}\beta},~~
\alpha, \beta>0 \label{prior-dists}
\end{equation}
The posterior distribution of $\theta$ via the Bayes rule is given by $\pi(\theta\mid \textbf{x}, \textbf{y})\propto \pi(\theta) L(\theta\mid \textbf{x}, \textbf{y})$, where $L(\theta\mid \textbf{x}, \textbf{y})$ is the likelihood function for $\theta$ based on $(\textbf{x}, \textbf{y})$, where its logarithm is given in (\ref{loglike}). The posterior distributions of
$\alpha$ and $\beta$ are independent and are given by
\[
\alpha|(\textbf{x}, \textbf{y})\sim Gamma(\gamma_{1}+n,\lambda_{1}-T_{1})
\]
\[
\beta|(\textbf{x}, \textbf{y})\sim Gamma(\gamma_{2}+m,\lambda_{2}-T_{2})
\]
where $T_{1} =\sum_{i=1}^{n}\log(1-e^{-(x_{i}+x_{i}^{2})})$ and
$T_{2} =\sum_{j=1}^{m}\log(1-e^{-(y_{j}+y_{j}^{2})})$.
\par
Bayesian inference on $R$ is based on the derivation of the posterior pdf of $R$, which can be obtained using a suitable one-to-one transformation of $\theta=(\alpha, \beta)$ of the form $G: \theta\to (R, \eta)$, with inverse $V=G^{-1}$, and $\eta=\alpha+\beta$. Then, the joint posterior pdf of $(R, \eta¸)$ is given by $\pi(R, \eta\mid \textbf{x, y})=\pi(V(R, \eta)\mid \textbf{x, y})|J_{V}(R,\eta)|$, where $|J_{V}(R,\eta)|$ is the Jacobian of the transformation $V$, so that
\[
\pi_{R}(r\mid \textbf{x, y}) =\int{\pi(V(r, \eta)\mid \textbf{x, y})|J_{V}(r,\eta)|d\eta}=\int{\pi(r, \eta\mid \textbf{x, y})|d\eta}
\]
Since a priori $\alpha$ and $\beta$ are independent, using the prior distributions presented in (\ref{prior-dists}), the joint posterior distribution of $(R, \eta)$
\[
\pi(r, \eta\mid \textbf{x, y})=C\eta^{\gamma_{1}+\gamma_{2}+n+m-1}\exp\{-\eta[r(\lambda_{1}-T_{1})-(1-r)(\lambda_{2}-T_{2})]\}r^{\gamma_{1}+n-1}(1-r)^{\gamma_{2}+m-1}
\]
where
\[
C=\frac{(\lambda_{1}-T_{1})^{\gamma_{1}+n}(\lambda_{2}-T_{2})^{\gamma_{2}+m}}{\Gamma(\gamma_{1}+n)\Gamma(\gamma_{2}+m)}
\]
Then, the marginal posterior density of $R$ is given by
\[
f_{R}(r\mid \textbf{x}, \textbf{y})=K\frac{r^{\gamma_{1}+n-1}(1-r)^{\gamma_{2}+m-1}}{[(\lambda_{1}-T_{1})r+(\lambda_{2}-T_{2})(1-r)]^{(n+m+\gamma_{1}+\gamma_{2})}}~\textrm{for}~~ 0<r<1
\]
where
\[
K=C\times\Gamma(n+m+\gamma_{1}+\gamma_{2})
\]
However, there is no close form for the posterior mean or median and the numerical method is required to derive them, but the posterior mode is the root of $\frac{d}{dr}f_{R}(r\mid \textbf{x, y}))=0$ and it is unique (see also Rezaei et al (2010) for the similar reasoning regarding the Generalized Pareto distribution).
\par
The Bayes estimate of $R$ under the squared error loss function, i.e., the posterior mean can be numerically obtained using the
numerical method presented in Lindley (1980) and Ahmad et al. (1997). This estimate of $R$ denoted by $\hat{R}_{B}$ is given by
\begin{equation}
\hat{R}_{B}=\tilde{R}[1+\frac{\tilde{\alpha}\tilde{R}^{2}(\tilde{\alpha}(n+\gamma_{1}-
1)-\tilde{\beta}(m
+\gamma_{2}-2))}{\tilde{\beta}^{2}(n+\lambda_{1}-1)(m+\lambda_{2}-1)}]\label{Bayes1},
\end{equation}
where $\tilde{R}=\frac{\tilde{\alpha}}{\tilde{\alpha}+\tilde{\beta}},
\tilde{\alpha}=\frac{n+\gamma_{1}-1}{\lambda_{1}-T_{1}}$ and
$\tilde{\beta}=\frac{m+\gamma_{2}-1}{\lambda_{2}-T_{2}}$.
\par
%
%
%
\section{Estimation of $R$ with common and unknown scale parameters}\label{unknown}

\subsection{Maximum likelihood estimator of $R$}
In this section, we wish to make inference about $R$ when the common scale parameters of $X$ and $Y$, that is, $(a, b)$ are unknown, and then investigate its properties. Let $(X_{1},X_{2},\ldots,X_{n})$ be a random sample from $GLFRD(a,b,\alpha)$ and $(Y_{1},Y_{2},\ldots,Y_{m})$ be a random sample from $GLFRD(a,b,\beta)$. To compute the MLE of $R$, the corresponding log-likelihood of the observed sample is given by
\[
\ell(a,b,\alpha,\beta)=n\ln\alpha+m\ln\beta+\sum_{i=1}^{n}\ln(a+bx_{i})+\sum_{j=1}^{m}\ln(a+by_{j})
\]
\[
+(\alpha-1)\sum_{i=1}^{n}\ln(1-e^{-(ax_{i}+\frac{b}{2}x_{i}^{2})})+(\beta-1)\sum_{j=1}^{m}\ln(1-e^{-(ay_{j}+\frac{b}{2}y_{j}^{2})})
\]
\[
-\sum_{i=1}^{n}(ax_{i}+\frac{b}{2}x_{i}^{2})-\sum_{j=1}^{m}(ay_{j}+\frac{b}{2}y_{j}^{2})\label{log-likelihood}
\]

The MLEs of $a,b,\alpha$ and $\beta$ say
$\hat{a},\hat{b},\hat{\alpha}$ and $\hat{\beta}$, respectively,
can be obtained as the solutions of the following equations

\[
\frac{\partial \ell}{\partial
a}=\sum_{i=1}^{n}\frac{1}{a+bx_{i}}+\sum_{j=1}^{m}\frac{1}{a+by_{j}}
\]
\begin{equation}
+(\alpha-1)\sum_{i=1}^{n}\frac{x_{i}e^{-(ax_{i}+\frac{b}{2}x_{i}^{2})}}
{1-e^{-(ax_{i}+\frac{b}{2}x_{i}^{2})}}+(\beta-1)\sum_{j=1}^{m}\frac{y_{j}
e^{-(ay_{j}+\frac{b}{2}y_{j}^{2})}}{1-e^{-(ay_{j}+\frac{b}{2}y_{j}^{2})}}
-\sum_{i=1}^{n}x_{i}-\sum_{j=1}^{m}y_{j},\label{MLEaa}
\end{equation}
\[
\frac{\partial \ell}{\partial
b}=\sum_{i=1}^{n}\frac{x_{i}}{a+bx_{i}}+\sum_{j=1}^{m}\frac{y_{j}}{a+by_{j}}
\]
\begin{equation}
+\frac{(\alpha-1)}{2}\sum_{i=1}^{n}\frac{x_{i}^{2}e^{-(ax_{i}+\frac{b}{2}x_{i}^{2})}}{1-e^{-(ax_{i}+\frac{b}{2}x_{i}^{2})}}+(\beta-1)\sum_{j=1}^{m}\frac{y_{j}^{2}e^{-(ay_{j}+\frac{b}{2}y_{j}^{2})}}{1-e^{-(ay_{j}+\frac{b}{2}y_{j}^{2})}}
-\sum_{i=1}^{n}\frac{x_{i}^{2}}{2}-\sum_{j=1}^{m}\frac{y_{j}^{2}}{2},\label{MLEbb}
\end{equation}
\begin{equation}
\frac{\partial \ell}{\partial
\alpha}=\frac{n}{\alpha}+\sum_{i=1}^{n}\ln(1-e^{-(ax_{i}+\frac{b}{2}x_{i}^{2})}),\label{MLEalpha}
\end{equation}
\begin{equation}
\frac{\partial \ell}{\partial
\beta}=\frac{m}{\beta}+\sum_{j=1}^{m}\ln(1-e^{-(ay_{j}+\frac{b}{2}y_{j}^{2})})\label{MLEbeta}
\end{equation}
From Equations (\ref{MLEalpha}) and (\ref{MLEbeta}), we can obtain the maximum likelihood estimates of $\alpha$ and $\beta$ as function of $a$ and $b$ as follows
\begin{equation}
\hat{\alpha}=\frac{-n}{\sum_{i=1}^{n}\ln(1-e^{-(ax_{i}+\frac{b}{2}x_{i}^{2})})}\label{hatalpha}
\end{equation}
and
\begin{equation}
\hat{\beta}=\frac{-m}{\sum_{j=1}^{m}\ln(1-e^{-(ay_{j}+\frac{b}{2}y_{j}^{2})})}\label{hatbeta}
\end{equation}
By replacing $\hat{\alpha},~\hat{\beta}$ in Equations (\ref{MLEaa}) and (\ref{MLEbb}), the MLEs of $a$ and $b$ can be then achieved as the solution of the following equations
\[
f_{1}(a,b\mid \hat{\alpha},~\hat{\beta})=\frac{\partial \ell}{\partial
a}=\sum_{i=1}^{n}\frac{1}{a+bx_{i}}+\sum_{j=1}^{m}\frac{1}{a+by_{j}}
\]
\[
+(\hat{\alpha}-1)\sum_{i=1}^{n}\frac{x_{i}e^{-(ax_{i}+\frac{b}{2}x_{i}^{2})}}{1-e^{-(ax_{i}+\frac{b}{2}x_{i}^{2})}}+(\hat{\beta}-1)\sum_{j=1}^{m}\frac{y_{j}e^{-(ay_{j}+\frac{b}{2}y_{j}^{2})}}{1-e^{-(ay_{j}+\frac{b}{2}y_{j}^{2})}}-\sum_{i=1}^{n}x_{i}-\sum_{j=1}^{m}y_{j},\label{MLEa}
\]
\[
f_{2}(a,b\mid \hat{\alpha},~\hat{\beta})=\frac{\partial \ell}{\partial
b}=\sum_{i=1}^{n}\frac{x_{i}}{a+bx_{i}}+\sum_{j=1}^{m}\frac{y_{j}}{a+by_{j}}
\]
\[
+\frac{(\hat{\alpha}-1)}{2}\sum_{i=1}^{n}\frac{x_{i}^{2}e^{-(ax_{i}+\frac{b}{2}x_{i}^{2})}}{1-e^{-(ax_{i}+\frac{b}{2}x_{i}^{2})}}+(\hat{\beta}-1)\sum_{j=1}^{m}\frac{y_{j}^{2}e^{-(ay_{j}+\frac{b}{2}y_{j}^{2})}}{1-e^{-(ay_{j}+\frac{b}{2}y_{j}^{2})}}-\sum_{i=1}^{n}\frac{x_{i}^{2}}{2}-\sum_{j=1}^{m}\frac{y_{j}^{2}}{2},\label{MLEb}
\]
As $\hat{a}, \hat{b}$ are the fixed points solution of the aforementioned equations, they can then be obtained by applying an iterative strategy as
\begin{eqnarray*}
f_{1}(a(i),b(i)\mid \hat{\alpha},~\hat{\beta})=0,~~~~~~f_{2}(a(i),b(i)\mid \hat{\alpha},~\hat{\beta})=0
\end{eqnarray*}
where $a(i), b(i)$ are the $i$th iteration of $\hat{a}, \hat{b}$.
\par
We should stop the iteration scheme when both $\|a(i+1)-a(i)\|$ and $\|b(i+1)-b(i)\|$ are adequately small. When $\hat{a}, \hat{b}$ are obtained, it would be straightforward to yield $\hat{\alpha}, \hat{\beta}$ from (\ref{hatalpha}) and (\ref{hatbeta}), respectively.
\par
Finally, due to the invariance property of the ML estimators, the MLE of $R$ will be as follows
\begin{equation}
\hat{R}=\frac{\hat{\alpha}}{\hat{\alpha}+\hat{\beta}}.\label{hatR}
\end{equation}
\subsection{Asymptotic distribution}

As the exact distribution of $\hat{R}$ does not exist, it is essential to investigate the asymptotic behaviour of the derived MLE of $R$ which is considered in this section. We first derive the asymptotic distribution of $\hat{\mbox{\boldmath{$\theta$}}}=(\hat{a},\hat{b},\hat{\alpha},\hat{\beta})$ and then the asymptotic distribution of $\hat{R}$ will be accordingly obtained. We then, based on the asymptotic distribution of $\hat{R}$, calculate the asymptotic confidence interval of $R$.
\par
We denote the observed information matrix of $\mbox{\boldmath{$\theta$}}=(a,b,\alpha,\beta)$ as $I=[I_{ij}]_{i,j=1,2,3,4} $, given by
 \begin{center}
\[
I(\mbox{\boldmath{$\theta$}})=\left(%
\begin{array}{cccc}
  \frac{\partial^{2}\ell}{\partial a^{2}} & \frac{\partial^{2}\ell}{\partial a\partial b} & \frac{\partial^{2}\ell}{\partial a\partial \alpha}& \frac{\partial^{2}\ell}{\partial a\partial\beta}\\
  \frac{\partial^{2}\ell}{\partial b\partial a} & \frac{\partial^{2}\ell}{\partial b^{2}} &\frac{\partial^{2}\ell}{\partial b\partial \alpha}& \frac{\partial^{2}\ell}{\partial b\partial \beta}\\
  \frac{\partial^{2}\ell}{\partial \alpha\partial a} &\frac{\partial^{2}\ell}{\partial \alpha\partial b} & \frac{\partial^{2}\ell}{\partial \alpha^{2}}&\frac{\partial^{2}\ell}{\partial \alpha\partial \beta}\\
  \frac{\partial^{2}\ell}{\partial\beta\partial a}&\frac{\partial^{2}\ell}{\partial\beta\partial b}&\frac{\partial^{2}\ell}{\partial\beta\partial\alpha}&\frac{\partial^{2}\ell}{\partial\beta^{2}}\\
\end{array}%
\right)
\]
\end{center}
where
\[
I_{11}=\sum_{i=1}^{n}\frac{1}{(a+bx_{i})^{2}}+(\alpha-1)\sum_{i=1}^{n}\frac{x_{i}^{2}e^{-(ax_{i}+\frac{b}{2}x_{i}^{2})}}{(1-e^{-(ax_{i}+\frac{b}{2}x_{i}^{2})})^{2}}
\]
\[
+\sum_{j=1}^{m}\frac{1}{(a+by_{j})^{2}}+(\beta-1)\sum_{j=1}^{m}\frac{y_{j}^{2}e^{-(ay_{j}+\frac{b}{2}y_{j}^{2})}}{(1-e^{-(ay_{j}+\frac{b}{2}y_{j}^{2})})^{2}}
\]
\[
I_{12}=I_{21}=\sum_{i=1}^{n}\frac{x_{i}}{(a+bx_{i})^{2}}+\sum_{j=1}^{m}\frac{y_{j}}{(a+by_{j})^{2}}+
\]
\[
\frac{(\alpha-1)}{2}\sum_{i=1}^{n}\frac{x_{i}^{3}e^{-(ax_{i}+\frac{b}{2}x_{i}^{2})}}{(1-e^{-(ax_{i}+\frac{b}{2}x_{i}^{2})})^{2}}
+\frac{(\beta-1)}{2}\sum_{j=1}^{m}\frac{y_{j}^{3}e^{-(ay_{j}+\frac{b}{2}y_{j}^{2})}}{(1-e^{-(ay_{j}+\frac{b}{2}y_{j}^{2})})^{2}}
\]
\[
I_{13}=I_{31}=-\sum_{i=1}^{n}\frac{x_{i}e^{-(ax_{i}+\frac{b}{2}x_{i}^{2})}}{1-e^{-(ax_{i}+\frac{b}{2}x_{i}^{2})}},~~
I_{14}=I_{41}=-\sum_{j=1}^{m}\frac{y_{j}e^{-(ay_{j}+\frac{b}{2}y_{j}^{2})}}{1-e^{-(ay_{j}+\frac{b}{2}y_{j}^{2})}}
\]
\[
I_{23}=I_{32}=-\sum_{i=1}^{n}\frac{x_{i}^{2}e^{-(ax_{i}+\frac{b}{2}x_{i}^{2})}}{2(1-e^{-(ax_{i}+\frac{b}{2}x_{i}^{2})})},~~I_{24}=I_{42}=-\sum_{j=1}^{m}\frac{y_{j}^{2}e^{-(ay_{j}+\frac{b}{2}y_{j}^{2})}}{2(1-e^{-(ay_{j}+\frac{b}{2}y_{j}^{2})})}
\]
\[
I_{22}=\sum_{i=1}^{n}\frac{x_{i}^{2}}{(a+bx_{i})^{2}}+\frac{(\alpha-1)}{4}\sum_{i=1}^{n}\frac{x_{i}^{4}e^{-(ax_{i}+\frac{b}{2}x_{i}^{2})}}{(1-e^{-(ax_{i}+\frac{b}{2}x_{i}^{2})})^{2}}
\]
\[
+\sum_{j=1}^{m}\frac{y_{j}^{2}}{(a+by_{j})^{2}}+\frac{(\beta-1)}{4}\sum_{j=1}^{m}\frac{y_{j}^{4}e^{-(ay_{j}+\frac{b}{2}y_{j}^{2})}}{(1-e^{-(ay_{j}+\frac{b}{2}y_{j}^{2})})^{2}}
\]
and
\[
I_{33}=\frac{n}{\alpha^{2}},~~I_{34}=I_{43}=0,~~I_{44}=\frac{m}{\beta^{2}}
\]

\textbf{Theorem 1.}~~  As $n, m\rightarrow\infty$ and $n/m \rightarrow p$ then
\[
[\sqrt{n}(\hat{a}-a),\sqrt{n}(\hat{b}-b),\sqrt{n}(\hat{\alpha}-\alpha),\sqrt{m}(\hat{\beta}-\beta)]\rightarrow N_{4}({\bf 0},{\bf U^{-1}}(\mbox{\boldmath{$\theta$}}))
\]
where
\begin{center}
\begin{equation}
{\bf U}(\mbox{\boldmath{$\theta$}})=\left(%
\begin{array}{cccc}
  u_{11} &u_{12}&u_{13}&u_{14}\\
  u_{21} &u_{22}&u_{23}&u_{24}\\
  u_{31} &u_{32}&u_{33}&0\\
  u_{41} &u_{42}&0&u_{44}\\
\end{array}%
\right)\label{U-equ}
\end{equation}
\end{center}
and
\[
u_{11}=\frac{1}{n}I_{11},~~~u_{12}=u_{21}=\frac{1}{n}I_{12},~~u_{13}=u_{31}=\frac{1}{n}I_{13},~~u_{14}=u_{41}=\frac{\sqrt{p}}{n}I_{14}
\]
\[
u_{22}=\frac{1}{n}I_{22},~~u_{23}=u_{32}=\frac{1}{n}I_{23},~~u_{24}=u_{42}=\frac{\sqrt{p}}{n}I_{24},~~u_{33}=\frac{1}{n}I_{33},~~u_{44}=\frac{1}{m}I_{44}
\]
\textbf{Proof.}~~ The proof follows from the asymptotic normality of
MLE (See Ferguson (1996) and references therein). $\blacksquare$\\
\par
\textbf{ Theorem 2.}~~As $n\rightarrow\infty$ and $m\rightarrow\infty$ and $n/m \rightarrow p$ then
$$\sqrt{n}(\hat{R}-R)\rightarrow N(\bf 0,{\sigma^{2}}),$$
where
\begin{equation}
{\sigma^{2}}=\frac{1}{k(\alpha+\beta)^{4}}[\beta^{2}a_{33}-2\sqrt{p}\alpha\beta a_{34}+\alpha^{2}{p}a_{44}]\label{Theorem2-sigma},
\end{equation}
\[
k=u_{11}u_{22}u_{33}u_{44}+u_{12}u_{23}u_{31}u_{44}+u_{12}u_{24}u_{33}u_{41}+u_{13}u_{21}u_{32}u_{44}+u_{13}u_{24}u_{31}u_{42}+
\]
\[
u_{14}u_{21}u_{33}u_{42}+u_{14}u_{23}u_{32}u_{41}-u_{11}u_{23}u_{32}u_{44}-u_{11}u_{24}u_{33}u_{42}-u_{12}u_{21}u_{33}u_{44}-
\]
\[
u_{13}u_{22}u_{31}u_{44}-u_{13}u_{24}u_{32}u_{41}-u_{14}u_{22}u_{33}u_{41}-u_{14}u_{23}u_{31}u_{42},
\]
\[
a_{33}=u_{11}u_{22}u_{44}+u_{12}u_{24}u_{41}+u_{14}u_{21}u_{42}-u_{11}u_{24}u_{42}-u_{12}u_{21}u_{44}-u_{14}u_{22}u_{41},
\]
\[
a_{34}=u_{11}u_{24}u_{32}+u_{14}u_{22}u_{31}-u_{12}u_{24}u_{31}-u_{14}u_{21}u_{32},\\
\]
\[
a_{44}=u_{11}u_{22}u_{33}+u_{12}u_{23}u_{31}+u_{13}u_{21}u_{32}-u_{11}u_{23}u_{32}-u_{12}u_{21}u_{33}-u_{13}u_{22}u_{31}
\]
\textit{Proof.}~~ See the Appendix.\\
\par
The motivation behind the asymptotic distribution presented above for $\hat{R}$ is to construct an asymptotic confidence interval for $R$. In order to construct this confidence interval, we first need to estimate ${\sigma^{2}}^{*}$. Duo to the invariance property of the ML estimator, we can estimate ${\sigma^{2}}^{*}$ by estimating its elements via replacing $({a},{b},{\alpha},{\beta})$ by their MLEs, $(\hat{a},\hat{b},\hat{\alpha},\hat{\beta})$. We will calculate this asymptotic confidence interval in Section 7 where the simulation results are presented.

\subsection{Confidence interval for Small sample size: Bootstrap approach}\label{bootsratp}

It would be reasonable to expect that the asymptotic confidence interval described above would not show satisfactory results when the sample size are small.  Efron (1982) proposes the \textit{percentile bootstrap method} (or \textit{Boot-p}) as an alternative way to construct a confidence interval in this situation.
\par
Algorithm of the percentile bootstrap method to estimate the confidence interval of $R$ is illustrated below:
\begin{enumerate}
\item From the sample $\{x_{1}, x_{2} ,\ldots, x_{n}\}$ and $\{y_{1},y_{2}, \ldots ,y_{m}\}$, compute
$\hat{a},\hat{b},\hat{\alpha}$ and $\hat{\beta}$.
\item Use $\hat{a},\hat{b}$ and $\hat{\alpha}$ to generate a bootstrap sample $\{x_{1}^{\ast},x_{2}^{\ast}, \ldots , x_{n}^{\ast}\}$ and similarly use $\hat{a},\hat{b}$ and $\hat{\beta}$  to generate a sample $\{y_{1}^{\ast},y_{2}^{\ast},\ldots,y_{m}^{\ast}\}$. Based on $\{x_{1}^{\ast}, x_{2}^{\ast},\ldots, x_{n}^{\ast}\}$ and $\{y_{1}^{\ast},y_{2}^{\ast},\ldots, y_{m}^{\ast}\}$ compute the bootstrap sample estimate of $R$ using
(\ref{hatR}), say $R^{\ast}$.
\item Repeat step 2, N boot times.
\item Let $G(x)=P(\hat{R}^{\ast}\leq x)$, be the cumulative distribution of
 $\hat{R}^{\ast}$. Define $R_{boot}^{\ast} =G^{-1}(x)$ for a given $x$. The approximate
 $100(1-\gamma)\%$ confidence interval of $R$ is given by
 \[
 (\hat{R}_{boot}(\frac{\gamma}{2}),\hat{R}_{boot}(1-\frac{\gamma}{2})).
 \]
\end{enumerate}
\subsection{Bayes Estimation of $R$}\label{Bayes2}
In this section, we derive the Bayes estimator of $R$. For constructing Bayes estimate of $R$, we assume independent Gamma priors on $a, b, \alpha$ and $\beta$ with the following pdfs
\begin{eqnarray*}
\pi_{1}(a)=\frac{\lambda_{1}^{\gamma_{1}}}{\Gamma(\gamma_{1})}a^{\gamma_{1}-1}e^{-\lambda_{1}a},~~~
\pi_{2}(b)=\frac{\lambda_{2}^{\gamma_{2}}}{\Gamma(\gamma_{2})}b^{\gamma_{2}-1}e^{-\lambda_{2}b},~~\\
\pi_{3}(\alpha)=\frac{\lambda_{3}^{\gamma_{3}}}{\Gamma(\gamma_{3})}\alpha^{\gamma_{3}-1}e^{-\lambda_{3}\alpha},~~~
\pi_{4}(\beta)=\frac{\lambda_{4}^{\gamma_{4}}}{\Gamma(\gamma_{4})}\beta^{\gamma_{4}-1}e^{-\lambda_{4}\beta},~~
a, b, \alpha, \beta>0
\end{eqnarray*}
where the hyper-parameters $\gamma_{i},\lambda_{i},~i=1,\ldots,4$ are all positive.
\par
The logarithm of posterior distribution of $a$ and $b$ after integrating out $\alpha$ and $\beta$ is as follows
\[
\log(\pi(a,b\mid \textbf{x,y}))=C+\sum_{i=1}^{n}\log(a+bx_{i})+\sum_{j=1}^{m}\log(a+by_{j})+\sum_{i=1}^{n}(ax_{i}+bx_{i}^{2})+\sum_{j=1}^{m}(ay_{j}+by_{j}^{2})
\]
\[
-(n+\gamma_{3})\log(\lambda_{3}+\sum_{i=1}^{n}(ax_{i}+bx_{i}^{2}))-(n+\gamma_{4})\log(\lambda_{4}+\sum_{j=1}^{m}(ay_{j}+by_{j}^{2}))
\]
\[
+(\gamma_{1}-1)\log(a)-\lambda_{1}a+(\gamma_{2}-1)\log(b)-\lambda_{2}b
\]
where $C$ is the normalizing constant.
\par
We estimate $a$ and $b$ by maximising $\log(\pi(a,b\mid \textbf{x,y}))$, that is,
\[
(\hat{a}, \hat{b})=\arg\max_{a,b}\log\{\pi(a,b\mid \textbf{x,y})\}
\]
where $\hat{a}$ and $\hat{b}$ are well-known as the maximum a posteriori (MAP) estimates of $a$ and $b$, respectively.
\par
We then substitute $(a, b)$ by $(\hat{a},\hat{b})$ in $\pi(\alpha\mid \textbf{x, y}, \hat{a},\hat{b})$ and $\pi(\beta\mid \textbf{x, y}, \hat{a},\hat{b})$ which are called \textit{pseudo-posteriors}. Since the parameters $\alpha$ and $\beta$ are assumed to be a priori independent, the posterior distributions of $\alpha$ and $\beta$ are then give by
\[
\alpha|(\textbf{x}, \textbf{y}, \hat{a},\hat{b})\sim Gamma(\gamma_{3}+n,\lambda_{3}-U_{1})
\]
\[
\beta|(\textbf{x}, \textbf{y}, \hat{a},\hat{b})\sim Gamma(\gamma_{4}+m,\lambda_{4}-U_{2})
\]
where $U_{1} =\sum_{i=1}^{n}\log(1-e^{-(\hat{a}x_{i}+\hat{b}x_{i}^{2})})$ and
$U_{2} =\sum_{j=1}^{m}\log(1-e^{-(\hat{a}y_{j}+\hat{b}y_{j}^{2})})$.
Similar to the method described in Subsection 3.2, using the prior distributions presented above, the marginal posterior density of $R$ becomes
\[
\pi_{R}(r\mid \textbf{x}, \textbf{y})=K_{1}\frac{r^{\gamma_{3}+n-1}(1-r)^{\gamma_{4}+m-1}}{[(\lambda_{3}-U_{1})r+(\lambda_{4}-U_{2})(1-r)]^{(n+m+\gamma_{3}+\gamma_{4})}},~~~ 0<r<1
\]
where
\[
K_{1}=\frac{(\lambda_{3}-U_{1})^{\gamma_{3}+n}(\lambda_{4}-U_{2})^{\gamma_{4}+m}\Gamma(n+m+\gamma_{3}+\gamma_{4})}{\Gamma(\gamma_{3}+n)\Gamma(\gamma_{4}+m)}.
\]
Although the Bayes estimates of $R$ under the squared error or absolute error loss function cannot be explicitly obtained (the numerical methods, such as, MCMC approach should be used), but the posterior mode could be derived in an explicit form. The derivative of $\pi_{R}(r\mid \textbf{x}, \textbf{y})$ can be easily calculated as follows:\\
\[
\frac{d\pi_{R}(r\mid \textbf{x}, \textbf{y})}{dr}=K_{1}H(r)h_{1}(r)
\]
where\\
\par
$h_{1}(r)=\frac{r^{\gamma_{3}+n-2}(1-r)^{\gamma_{4}+m-2}}{[(\lambda_{3}-U_{1})r+(\lambda_{4}-U_{2})(1-r)]^{(n+m+\gamma_{3}+\gamma_{4}+1)}}$\\
\[
H(r)=\{[A_{1}(1-r)-A_{2}r][(\lambda_{3}-U_{1})r+(\lambda_{4}-U_{2})(1-r)]
-A_{3}[(\lambda_{3}-U_{1})-(\lambda_{4}-U_{2})]r(1-r)\}
\]
and
\[
A_{1}=\gamma_{3}+n-1,~~~A_{2}=\gamma_{4}+m-1,~~~~~A_{3}=n+m+\gamma_{3}+\gamma_{4}
\]

Depending on the signs of $(\lambda_{4}-U_{2})$ and $(\lambda_{3}-U_{1})$, one can easily show that $\pi_{R}(r\mid \textbf{x}, \textbf{y})$ has a unique mode over $0 < r < 1$, and the corresponding posterior mode can be obtained as the unique root of the equation $H(r) = 0$ over $0 < r < 1$.
\par
A reasonable loss function to estimate $R$ is
\begin{equation}
L(d_{1},d_{2})=\left\{ \begin{array}{ll}
0 & ~~~\mbox{if~~ $|d_{1}-d_{2}|\leq c$};\\
1 & ~~~\mbox{if~~ $|d_{1}-d_{2}|>c$}.
\end{array} \right.\label{01Loss}
\end{equation}
Fergusen (1967) computed the Bayes estimate under the loss function given in (\ref{01Loss}), as he
midpoint of the modal interval of length $2c$ of the posterior distribution. As a result, the posterior mode can be considered as an approximate Bayes estimate of $R$ with respect to the loss function presented in (\ref{01Loss}) when the constant $c$ is small. The credible interval of $R$ can be obtained by using numerical integration. For example, Chen and Shao (1999) introduce a MCMC method to compute the highest
posterior density (HPD) interval which will not be addressed here.
\par
An alternative simulation method that can be used here is originally proposed by Devroye (1984) and then used by Sara\c{c}o\v{g}lu et al (2012) to generate a sample from the posterior density function of $R$. It is then trivial to compute the Bayes estimate of $R$ and the associated credible interval based on this sample. Since the support of $\pi_{R}(r\mid \textbf{x}, \textbf{y})$, that is, $0<r<1$ is bounded, we use the acceptance rejection method to simulate a sample from $\pi_{R}(r\mid \textbf{x}, \textbf{y})$. Therefore, in order to compute the Bayes estimate and the credible interval, we implement the following steps. First, we determine the posterior mode of $\pi_{R}(r\mid \textbf{x}, \textbf{y})$, denoted by $\hat{r}_{M}$, as explained above. Therefore, for any $0<r<1$, it can be concluded that $\pi_{R}(r\mid \textbf{x}, \textbf{y})\leq \pi_{R}(\hat{r}_{M}\mid \textbf{x}, \textbf{y})$. Using acceptance rejection method introduced by Devroye (1984), we generate $l$ samples, denoted by $r_{1},r_{2},\ldots, r_{l}$, from $\pi_{R}(r\mid \textbf{x}, \textbf{y})$ given in (34). Now, the Bayes estimate of $R$ with respect to the squared error loss function can be calculated as the sample mean. The lower and upper $\alpha/2$-th percentile points of the ordered sample can be served as the lower and upper bounds of 100$(1-\alpha)\%$ confidence interval, respectively.
\par
The Bayes estimate of $R$ under the squared error loss function can also be obtained using the numerical method studied in Lindley (1980) and Ahmad et al. (1997) and as presented in Section 3.2. The Bayes estimate of $R$, denoted by $\hat{R}_{B}$, is given by
\[
\hat{R}_{B}=\tilde{R}[1+\frac{\tilde{\alpha}\tilde{R}^{2}(\tilde{\alpha}(n+\gamma_{3}-
1)-\tilde{\beta}(m
+\gamma_{4}-2))}{\tilde{\beta}^{2}(n+\lambda_{3}-1)(m+\lambda_{4}-1)}],
\]
where $\tilde{R}=\frac{\tilde{\alpha}}{\tilde{\alpha}+\tilde{\beta}},
\tilde{\alpha}=\frac{n+\gamma_{3}-1}{\lambda_{3}-U_{1}}$ and
$\tilde{\beta}=\frac{m+\gamma_{4}-1}{\lambda_{4}-U_{2}}$.
\par
%
%

\section{Estimation of $R$ in the general case}\label{general}
In this section, we present the estimations of the stress-strength model, $R=P(Y<X)$, when $X\sim GLFR(a_{1},b_{1},\alpha )$ and $Y\sim GLFR(a_{2},b_{2},\beta)$. We present the MLE of $R$ and its associated confidence intervals in the next subsection. We also present the Bayes estimate of $R$ later in this section.
\subsection{Maximum likelihood estimator of R}
Let $X\sim GLFR(a_{1},b_{1},\alpha )$ and $Y\sim GLFR(a_{2},b_{2},\beta)$, and these two random variables are assumed to be independent. Therefore,
\begin{eqnarray*}
R = P [Y < X ] = \int_{0}^{\infty} P( Y < X |X = x
)f_{X}(x)dx\\
=\int_{0}^{\infty}\alpha(a_{1}+b_{1}x)e^{-(a_{1}x+\frac{b_{1}}{2}x^{2})}(1-e^{-(a_{1}x+\frac{b_{1}}{2}x^{2})})^{\alpha-1}(1-e^{-(a_{2}x+\frac{b_{2}}{2}x^{2})})^{\beta}dx\\
\end{eqnarray*}
Suppose further $X_1 , X_2 , \ldots , X_n$ is a random sample from
$GLFRD(a_{1},b_{1},\alpha)$ and $Y_1 , Y_2 , \ldots , Y_m$ is another random sample from $GLFRD(a_{2},b_{2},\beta)$. The log-likelihood function of the observed samples is presented as
\begin{eqnarray*}
\ell(a_{1},b_{1},a_{2},b_{2},\alpha,\beta)=n\ln\alpha+m\ln\beta+\sum_{i=1}^{n}\ln(a_{1}+b_{1}x_{i})+\sum_{j=1}^{m}\ln(a_{2}+b_{2}y_{j})\\
                   +(\alpha-1)\sum_{i=1}^{n}\ln(1-e^{-(a_{1}x_{i}+\frac{b_{1}}{2}x_{i}^{2})})+(\beta-1)\sum_{j=1}^{m}\ln(1-e^{-(a_{2}y_{j}+\frac{b_{2}}{2}y_{j}^{2})})\\
                   -\sum_{i=1}^{n}(a_{1}x_{i}+\frac{b_{1}}{2}x_{i}^{2})-\sum_{j=1}^{m}(a_{2}y_{j}+\frac{b_{2}}{2}y_{j}^{2})\\
\end{eqnarray*}

The MLEs of $a_{1},b_{1},a_{2},b_{2},\alpha$ and $\beta$ say
$\hat{a_{1}},\hat{b_{1}},\hat{a_{2}},\hat{b_{2}},\hat{\alpha}$ and $\hat{\beta}$ , respectively, can be obtained
as the solutions of
\begin{equation}
\frac{\partial \ell}{\partial
a_{1}}=\sum_{i=1}^{n}\frac{1}{a_{1}+b_{1}x_{i}}+(\alpha-1)\sum_{i=1}^{n}\frac{x_{i}e^{-(a_{1}x_{i}+\frac{b_{1}}{2}x_{i}^{2})}}{1-e^{-(a_{1}x_{i}+\frac{b_{1}}{2}x_{i}^{2})}}
-\sum_{i=1}^{n}x_{i}\label{MLEa1}
\end{equation}
\begin{equation}
\frac{\partial \ell}{\partial
a_{2}}=\sum_{j=1}^{m}\frac{1}{a_{2}+b_{2}y_{j}}+(\beta-1)\sum_{j=1}^{m}\frac{y_{j}e^{-(a_{2}y_{j}
+\frac{b_{2}}{2}y_{j}^{2})}}{1-e^{-(a_{2}y_{j}+\frac{b_{2}}{2}y_{j}^{2})}}-\sum_{j=1}^{m}y_{j}\label{MLEa2}
\end{equation}
\begin{equation}
\frac{\partial \ell}{\partial
b_{1}}=\sum_{i=1}^{n}\frac{x_{i}}{a_{1}+b_{1}x_{i}}+\frac{(\alpha-1)}{2}
\sum_{i=1}^{n}\frac{x_{i}^{2}e^{-(a_{1}x_{i}+\frac{b_{1}}{2}x_{i}^{2})}}{1-e^{-(a_{1}x_{i}+\frac{b_{1}}{2}x_{i}^{2})}}
-\sum_{i=1}^{n}\frac{x_{i}^{2}}{2}\label{MLEb1}
\end{equation}
\begin{equation}
\frac{\partial \ell}{\partial
b_{2}}=\sum_{j=1}^{m}\frac{y_{j}}{a_{2}+b_{2}y_{j}}+(\beta-1)
\sum_{j=1}^{m}\frac{y_{j}^{2}e^{-(a_{2}y_{j}+\frac{b_{2}}{2}y_{j}^{2})}}
{1-e^{-(a_{2}y_{j}+\frac{b_{2}}{2}y_{j}^{2})}}-\sum_{j=1}^{m}\frac{y_{j}^{2}}{2}\label{MLEb2}
\end{equation}
\begin{equation}
\frac{\partial \ell}{\partial
\alpha}=\frac{n}{\alpha}+\sum_{i=1}^{n}\ln(1-e^{-(a_{1}x_{i}+\frac{b_{1}}{2}x_{i}^{2})})\label{alphaG}
\end{equation}
\begin{equation}
\frac{\partial \ell}{\partial
\beta}=\frac{m}{\beta}+\sum_{j=1}^{m}\ln(1-e^{-(a_{2}y_{j}+\frac{b_{2}}{2}y_{j}^{2})}) \label{betaG}
\end{equation}

Similarly, from Equations (\ref{alphaG}) and (\ref{betaG}), the MLEs of $\alpha$ and $\beta$ can be obtained, as a function of
$a_{1}, a_{2}, b_{1}, b_{2}$, as follows
\begin{equation}
\hat{\alpha}=\frac{-n}{\sum_{i=1}^{n}\ln(1-e^{-(a_{1}x_{i}+\frac{b_{1}}{2}x_{i}^{2})})}\label{hatalphaG}
\end{equation}
and
\begin{equation}
\hat{\beta}=\frac{-m}{\sum_{j=1}^{m}\ln(1-e^{-(a_{2}y_{j}+\frac{b_{2}}{2}y_{j}^{2})})}\label{hatbetaG}
\end{equation}
Then, the MLEs of $a_{1},b_{1}$ denoted by $\hat{a_{1}},\hat{b_{1}}$ can be obtained by substituting $\hat{\alpha}$ in Equations (\ref{MLEa1}),(\ref{MLEb1}), as the solution of the following equations
\[
g_{1}(a_{1},b_{1}\mid \hat{\alpha})=
\sum_{i=1}^{n}\frac{1}{a_{1}+b_{1}x_{i}}+(\hat{\alpha}-1)\sum_{i=1}^{n}\frac{x_{i}e^{-(a_{1}x_{i}+\frac{b_{1}}{2}x_{i}^{2})}}{1-e^{-(a_{1}x_{i}+\frac{b_{1}}{2}x_{i}^{2})}}
-\sum_{i=1}^{n}x_{i}\label{EQMLEa1}
\]
\[
g_{2}(a_{1},b_{1}\mid \hat{\alpha})=\sum_{i=1}^{n}\frac{x_{i}}{a_{1}+b_{1}x_{i}}+\frac{(\hat{\alpha}-1)}{2}
\sum_{i=1}^{n}\frac{x_{i}^{2}e^{-(a_{1}x_{i}+\frac{b_{1}}{2}x_{i}^{2})}}{1-e^{-(a_{1}x_{i}+\frac{b_{1}}{2}x_{i}^{2})}}
-\sum_{i=1}^{n}\frac{x_{i}^{2}}{2}\label{WQMLEb1}
\]
\[
+\frac{(\hat{\alpha}-1)}{2}\sum_{i=1}^{n}\frac{x_{i}^{2}e^{-(ax_{i}+\frac{b}{2}x_{i}^{2})}}{1-e^{-(ax_{i}+\frac{b}{2}x_{i}^{2})}}+(\hat{\beta}-1)\sum_{j=1}^{m}\frac{y_{j}^{2}e^{-(ay_{j}+\frac{b}{2}y_{j}^{2})}}{1-e^{-(ay_{j}+\frac{b}{2}y_{j}^{2})}}-\sum_{i=1}^{n}\frac{x_{i}^{2}}{2}-\sum_{j=1}^{m}\frac{y_{j}^{2}}{2},\label{MLEb}
\]
In a similar way, the MLEs of $a_{2},b_{2}$ denoted by $\hat{a_{2}},\hat{b_{2}}$ can be obtained by substituting $\hat{\beta}$ in Equations (\ref{MLEa2}),(\ref{MLEb2}), as the solution of the following equations
\[
g_{3}(a_{2},b_{2}\mid \hat{\beta})=\sum_{j=1}^{m}\frac{1}{a_{2}+b_{2}y_{j}}+(\hat{\beta}-1)\sum_{j=1}^{m}\frac{y_{j}e^{-(a_{2}y_{j}
+\frac{b_{2}}{2}y_{j}^{2})}}{1-e^{-(a_{2}y_{j}+\frac{b_{2}}{2}y_{j}^{2})}}-\sum_{j=1}^{m}y_{j}\label{EQMLEa2}
\]
\[
g_{4}(a_{2},b_{2}\mid \hat{\beta})=\sum_{j=1}^{m}\frac{y_{j}}{a_{2}+b_{2}y_{j}}+(\hat{\beta}-1)
\sum_{j=1}^{m}\frac{y_{j}^{2}e^{-(a_{2}y_{j}+\frac{b_{2}}{2}y_{j}^{2})}}
{1-e^{-(a_{2}y_{j}+\frac{b_{2}}{2}y_{j}^{2})}}-\sum_{j=1}^{m}\frac{y_{j}^{2}}{2}\label{EQMLEb2}
\]
Knowing that $\hat{a}_{1}, \hat{a}_{2}, \hat{b}_{1}, \hat{b}_{2}$ are the fixed points solution of the aforementioned equations, we can then drive them by applying the similar iterative scheme, used in the previous section, as
\begin{eqnarray*}
g_{1}(a_{1}(i),b_{1}(i)\mid \hat{\alpha})=0,~~~~~~g_{2}(a_{1}(i),b_{1}(i)\mid \hat{\alpha})=0
\end{eqnarray*}
\begin{eqnarray*}
g_{3}(a_{2}(i),b_{2}(i)\mid \hat{\beta})=0,~~~~~~g_{4}(a_{2}(i),b_{2}(i)\mid \hat{\beta})=0
\end{eqnarray*}
where $a_{1}(i),a_{2}(i), b_{1}(i), b_{2}(i)$ are the $i$th iteration corresponding to $\hat{a}_{1}, \hat{a}_{2}, \hat{b}_{1}, \hat{b}_{2}$.
\par
We will stop this iteration scheme when $\|a_{j}(i+1)-a_{j}(i)\|$ and $\|b_{j}(i+1)-b_{j}(i)\|$, $j=1,2$, are adequately small. When $\hat{a}_{j}, \hat{b}_{j},~j=1,2$ are obtained, it would be straightforward to calculate $\hat{\alpha}, \hat{\beta}$ from (\ref{hatalphaG}) and (\ref{hatbetaG}), respectively.
\par
Finally, due to the invariance property of the ML estimators, the MLE of $R$ will be as follows
\[
\hat{R}= \int_{0}^{\infty}\hat{\alpha}(\hat{a}_{1}+\hat{b}_{1}x)e^{-(\hat{a}_{1}x+\frac{\hat{b}_{1}}{2}x^{2})}(1-e^{-(\hat{a}_{1}x+\frac{\hat{b}_{1}}{2}x^{2})})^{\hat{\alpha}-1}(1-e^{-(\hat{a}_{2}x+\frac{\hat{b}_{2}}{2}x^{2})})^{\hat{\beta}}dx\\
\label{hatRG}
\]
\subsection{Bayes Estimation of $R$}
To construct a Bayes estimate for $R$, we consider the following Gamma prior distributions on  $a_{1},a_{2}, b_{1}, b_{2} \alpha$ and $\beta$ as follows
\begin{eqnarray*}
a_{1}\sim Gamma(\lambda_{1},\gamma_{1}),~~b_{1}\sim Gamma(\lambda_{2},\gamma_{2}),~~a_{2}\sim Gamma(\lambda_{3},\gamma_{3})\\
b_{2}\sim Gamma(\lambda_{4},\gamma_{4}),~~\alpha\sim Gamma(\lambda_{5},\gamma_{5}),~~\beta\sim Gamma(\lambda_{6},\gamma_{6})
\end{eqnarray*}
Furthermore, we assume all these parameters to be independent of each other.
\par
Similar to the approach used in Subsection 4.4, we first estimate $(a_{1},b_{1},a_{2}, b_{2})$, denoted by $(\hat{a}_{1},\hat{b}_{1},\hat{a}_{2}, \hat{b}_{2})$,  by maximizing the associated posterior distribution as follows
\[
\log(\pi(a_{1},b_{1},a_{2}, b_{2}\mid \textbf{x,y}))=C+\sum_{i=1}^{n}\log(a_{1}+b_{1}x_{i})+\sum_{j=1}^{m}\log(a_{2}+b_{2}y_{j})+\sum_{i=1}^{n}(a_{1}x_{i}+b_{1}x_{i}^{2})
\]
\[
+\sum_{j=1}^{m}(a_{2}y_{j}+b_{2}y_{j}^{2})-(n+\gamma_{3})\log(\lambda_{3}+\sum_{i=1}^{n}(a_{1}x_{i}+bx_{i}^{2}))-(n+\gamma_{4})\log(\lambda_{4}+\sum_{j=1}^{m}(a_{2}y_{j}+b_{2}y_{j}^{2}))+
\]
\[
[(\gamma_{1}-1)\log(a_{1})-\lambda_{1}a_{1}]+[(\gamma_{3}-1)\log(b_{1})-\lambda_{3}b_{1}]+[(\gamma_{2}-1)\log(a_{2})-\lambda_{2}a_{2}]+[(\gamma_{4}-1)\log(b_{2})-\lambda_{4}b_{2}]
\]
We then substitute these estimates in $\pi(\alpha, \beta\mid \textbf{x, y}, \hat{a}_{1},\hat{b}_{1},\hat{a}_{2}, \hat{b}_{2})$ which is called a pseudo-posterior. As a result, the posterior distributions of $\alpha$ and $\beta$ are give by
\[
\alpha|(\textbf{x}, \textbf{y}, \hat{a}_{1},\hat{b}_{1},\hat{a}_{2}, \hat{b}_{2})\sim Gamma(\gamma_{5}+n,\lambda_{5}-V_{1})
\]
\[
\beta|(\textbf{x}, \textbf{y}, \hat{a}_{1},\hat{b}_{1},\hat{a}_{2}, \hat{b}_{2})\sim Gamma(\gamma_{6}+m,\lambda_{6}-V_{2})
\]
where $V_{1} =\sum_{i=1}^{n}\log(1-e^{-(\hat{a}_{1}x_{i}+\hat{b}_{1}x_{i}^{2})})$ and
$V_{2} =\sum_{j=1}^{m}\log(1-e^{-(\hat{a}_{2}y_{j}+\hat{b}_{2}y_{j}^{2})})$.
\par
It is then trivial (as shown in Subsection 3.2) to calculate the Bayes estimate of $R$ under the squared error loss function as follows
\[
\hat{R}_{B}=\tilde{R}[1+\frac{\tilde{\alpha}\tilde{R}^{2}(\tilde{\alpha}(n+\gamma_{5}-
1)-\tilde{\beta}(m
+\gamma_{6}-2))}{\tilde{\beta}^{2}(n+\lambda_{5}-1)(m+\lambda_{6}-1)}],
\]
where $\tilde{R}=\frac{\tilde{\alpha}}{\tilde{\alpha}+\tilde{\beta}},
\tilde{\alpha}=\frac{n+\gamma_{5}-1}{\lambda_{5}-V_{1}}$ and
$\tilde{\beta}=\frac{m+\gamma_{6}-1}{\lambda_{6}-V_{2}}$.
\par
An alternative way to estimate the Bayes estimate of $R$ under the squared error loss function is the simulation method described in Subsection 4.4 at which a sample from the posterior density function of $R$ can be generated using the acceptance rejection sampling method proposed by Devroye (1984). The mean of this sample can be served as the Bayes estimate of $R$, and the lower and upper $\alpha/2$-th percentile points of the ordered generated sample can then be considered as the lower and upper bounds of 100$(1-\alpha)\%$ credible interval, respectively.

\section{Estimation of $R$ Using progressively Censored Sample}

The main objective of this section it to address the the statistical inference of the stress-strength parameter $R =
P(X < Y )$ when $X$ and $Y$ are independent generalized linear failure rate random variables. It is further assumed that we observe progressively \textit{type-II censored} samples from $X\sim GLFRD(a,b,\alpha)$ and $Y\sim GLFRD(a,b,\beta)$ under different censoring schemes. We only consider the MLE of $R$, and the Bayesian inference about $R$ is being presented in a working paper by the authors. The Bayesian methods used to estimate $R$ in this case are very similar to the methods used by Kim and Chung (2006) to estimate $P (Y < X)$ when $X$ and $Y$ are both distributed as Burr-type $X$ model.
\par
Although there have been extensive works regarding the developments of the stress-strength
models under complete samples, but not much attention has been paid to the case at which the data are censored. Jiang and Wong (2008) and Sara\c{c}o\u{g}lu et al (2012) are among the pioneering works which report the estimation of $R$ for the exponential distribution under some censoring schemes. We are going to adopt Sara\c{c}o\u{g}lu et al's work to estimate  the stress-strength parameter for the generalized linear failure rate distributions under
progressive type-II censoring sampling schemes.
\par
We first briefly explain the progressive type-II censoring scheme, and then we derive the MLE of $R$ when the observed data are progressively type-II censored samples from these distributions.
\par
In medical or industrial applications, researchers have to treat the censored data because
they usually do not have sufficient time to observe the lifetime of all subjects in the study.
Furthermore, subjects/items may fail by cause other than the ones under study. There
are numerous schemes of censoring. There are several types of censoring: Type I and II censoring; random censoring (including right and left censoring); interval censoring; and truncation. Among these censoring schemes, the first two, type-I and type-II, are the two most well-known censoring
schemes. While in type-I censoring scheme, the experiment is stopped at a pre-fixed time point, in type-II censoring scheme, the experiment is stopped whenever a fixed number of failures (pre-fixed) has been observed. Sara\c{c}o\u{g}lu et al (2012) combine the type-II censoring and progressive censoring schemes which is called the \textit{progressive type-II censoring}. This scheme allows the researcher to remove active units during the experiment and is defined as follows: Given $m < n$, and $P_{1},\ldots,P_{m}$ non-negative integers such that
\[
P_{1} + \ldots + P_{m} = n -m
\]
where $n$ stands for the items are on the life test at the same time. At the time of the first failure, one chooses randomly $P_{1}$ items from the rest of the active $n-1$ and then discards. In the similar way, at time of the second failure, one selects $P_{2}$ out of $n-P_{1}-2$ remaining items at random and consequently remove it, and so on. Eventually, at the time of the $m$-th failure, all the remaining active items
are removed (see Sara\c{c}o\u{g}lu et al (2012) and reference therein for the advantages of this censoring scheme).
\par
We consider two progressive censoring schemes, namely $\{n_{1},m_{1},P_{1},P_{2},\ldots,P_{m_{1}}\}$ and $\{n_{2},m_{2}, Q_{1}, Q_{2},\ldots, Q_{m_{2}}\}$ for $X\sim GLFRD(a,b,\alpha)$ and $Y\sim GLFRD(a,b,\beta)$, respectively. The progressively censored
samples are observed as $\textbf{X}=(X_{1: m_{1}:
n_{1}},\ldots,X_{m_{1}: m_{1}: n_{1}})$ from $X$ and $\textbf{Y}=(Y_{1: m_{2}: n_{2}},\ldots,Y{m_{2}: m_{2}: n_{2}})$ from $Y$.
\par
The joint pdf of $\textbf{X}=(X_{1: m_{1}: n_{1}},\ldots,X_{m_{1}: m_{1}: n_{1}})$ is
\[
f_{X_{1: m_{1}: n_{1}},\ldots,X_{m_{1}: m_{1}:
n_{1}}}(x_{1},\ldots,x_{m_{1}})=c
\alpha^{m_{1}}\exp\{-\sum_{i=1}^{m_{1}}(ax_{i:m_{1}:n_{1}}+\frac{b}{2}x_{i:m_{1}:n_{1}}^{2})\}\times
\]
\[
\prod_{i=1}^{m_{1}}[(a+bx_{i:m_{1}:n_{1}})
(1-\exp\{-(ax_{i:m_{1}:n_{1}}+\frac{b}{2}x_{i:m_{1}:n_{1}}^{2})\})^{\alpha-1}
(1-(1-\exp\{-(ax_{i:m_{1}:n_{1}}+\frac{b}{2}x_{i:m_{1}:n_{1}}^{2})\})^{\alpha})^{P_{i}}]\label{X-samples}
\]
where
$c=n_{1}(n_{1}-P_{1}-1)\ldots(n_{1}-P_{1}-\ldots-P_{m_{1}-1}-m_{1}+1)$ is
the normalizing constant (Balakrishnan and Aggarwala, 2000).
\par
Similarly, the joint pdf of $\textbf{Y}$ can be obtained by replacing the $\textbf{X}$ values, $m_{1}$,  $n_{1}$ and $\{P_{1},\ldots,P_{m_{1}}\}$ by the $\textbf{Y}$ values, $m_{2}$,  $n_{2}$ and $\{Q_{1},\ldots,Q_{m_{2}}\}$, respectively.
\par
Therefore, the log-likelihood function of the progressively
censored sample is
\begin{eqnarray*}
\ell(a,b,\alpha,\beta)=m_{1}\ln\alpha+m_{2}\ln\beta+\sum_{i=1}^{m_{1}}\ln(a+bx_{i})+
\sum_{j=1}^{m_{2}}\ln(a+by_{j})\\
                   +(\alpha-1)\sum_{i=1}^{m_{1}}\ln(1-e^{-(ax_{i}+\frac{b}{2}x_{i}^{2})})+
                   (\beta-1)\sum_{j=1}^{m_{2}}\ln(1-e^{-(ay_{j}+\frac{b}{2}y_{j}^{2})})\\
                   +\sum_{i=1}^{m_{1}}P_{i}\ln(1-(1-e^{-(ax_{i}+\frac{b}{2}x_{i}^{2})})^{\alpha})-\sum_{i=1}^{m_{1}}(ax_{i}+\frac{b}{2}x_{i}^{2})\\
                   +\sum_{j=1}^{m_{2}}Q_{j}\ln(1-(1-e^{-(ay_{j}+\frac{b}{2}y_{j}^{2})})^{\beta})-\sum_{j=1}^{m_{2}}(ay_{j}+\frac{b}{2}y_{j}^{2})\\
\end{eqnarray*}
The MLEs of $\alpha$ and $\beta$, denoted by $\hat{\alpha}$ and
$\hat{\beta}$ , respectively, can be obtained as the solutions of the following equations by the iterative scheme described in the previous sections
\[
\frac{\partial \ell}{\partial
\alpha}=\frac{m_{1}}{\alpha}+\sum_{i=1}^{m_{1}}\ln(1-e^{-(ax_{i}+\frac{b}{2}x_{i}^{2})})-\sum_{i=1}^{m_{1}}
P_{i}\frac{(1-e^{-(ax_{i}+\frac{b}{2}x_{i}^{2})})^{\alpha}\ln(1-e^{-(ax_{i}+\frac{b}{2}x_{i}^{2})})}{(1-(1-e^{-(ax_{i}+\frac{b}{2}x_{i}^{2})})^{\alpha})}
\]
\[
\frac{\partial \ell}{\partial
\beta}=\frac{m_{2}}{\beta}+\sum_{j=1}^{m_{2}}\ln(1-e^{-(ay_{j}+\frac{b}{2}y_{j}^{2})})-\sum_{j=1}^{m_{2}}
Q_{j}\frac{(1-e^{-(ay_{j}+\frac{b}{2}y_{j}^{2})})^{\beta}\ln(1-e^{-(ay_{j}+\frac{b}{2}y_{j}^{2})})}{(1-(1-e^{-(ay_{j}+\frac{b}{2}y_{j}^{2})})^{\beta})}
\]
Finally, due to the invariance property of the ML estimators, the MLE of $R$ will be as follows
\[
\hat{R}=\frac{\hat{\alpha}}{\hat{\alpha}+\hat{\beta}}.\label{hatRG}
\]


\section{Simulation Results}
In this section, we present some results based on Monte Carlo simulations to compare the performances of the different estimators described in Sections 3 to 6, under the complete observed data, and also under progressive censored observed data.  We consider these two cases separately
to draw inference about $R$. We first assume that the data are complete and the common scale parameters $a,b$ are also known. In this case, we
consider combination of the small sample sizes: m, n = 15, 25 and 50. Without loss of generality, we set $a=1, b=2$. Table \ref{Tab1} illustrates the stress–-strength parameter, $R$, the MLE ($\hat{R}$), the Bayes estimate ($\hat{R}_{B}$), the confidence interval based on $\hat{R}$ denoted by $CI_{MLE}$, and its coverage percentage ($cp$), based on the simulated data from the GLFR distributions with the different values of $\alpha$ and $\beta$.
\par
The Bayes estimate of $R$ is computed, using (\ref{Bayes1}), with respect to the given Gamma prior distributions on $\alpha$ and $\beta$. It would be quite conventional to use the non-informative prior distributions for $\alpha$ and $\beta$. To avoid having the improper posterior distribution, we set the hyper-parameters of the Gamma distributions as $\gamma_{1}=\gamma_{2}=\lambda_{1}=\lambda_{2}=0.0001$ (see Kundu and Gupta, 2005). This is trivial to show that the bias and variance of the Bayes estimate would decrease as one could elicit a more informative prior distributions for $\alpha$ and $\beta$ (see O'Hagan et al. 2006).
\par
When the scale parameters $a, b$ are unknown, we first simulate data based on the mentioned sample sizes and the parameters illustrated in the first and second columns of Table \ref{tab2}. We then based on the methods described in Sections \ref{unknown} and \ref{general}, estimate $R$. Table \ref{tab2} also illustrates the average biases (Bias), mean square error (MSE) of $\hat{R}$, the confidence intervals (based on the asymptotic distribution of $\hat{R}$ and using the estimation of $\sigma^{2}$ given in (\ref{Theorem2-sigma})), and its associated coverage percentages ($cp$) based on 1000 replications.  Based on the reported biases and MSEs of ($\hat{R}$), one can see, even for the small sample sizes, the precisions of the calculated MLEs are quite satisfactory, and as expected the MSEs and biases become smaller as more sample sizes gathered which coincides with the consistency property of the MLE estimators.
\par
Similar to the findings in Kundu and Gupta (2005), Rezaei et al (2010) and Franc\'{e}s and Montoya (2012), the derived confidence intervals based on the MLEs perform very well unless the
sample size is quite small (e.g., when $m=n=15$). In addition, the coverage percentage of the asymptotic confidence interval will tend to the
nominal level, $95\%$ as the sample sizes increase.
\par
It is reasonable to expect that the confidence intervals approximated based on the asymptotic results for small sample sizes should not show satisfactory results. An alternative way is to use the bootstrap method explained in Subsection \ref{bootsratp}. In order to calculate these intervals, we first generate 20 numbers as a sample from  $GLFRD(1,0.4,1.5)$ and 20 other numbers as a second sample from $GLFRD(1,0.4,1)$ given in Table \ref{tab3}. We then start analyzing these data by calculating the MLEs of $(a, b,\alpha,\beta)$ using the iterative procedures described above. They are reported in the second column of Table \ref{tab4} for two cases: the common scale parameters are known or unknown. We use the percentile bootstrap method described in Subsection \ref{bootsratp} to compute the confidence interval for these data. In Table \ref{tab4}, $\hat{R}^{*}$ and $CI_{boot}$ denote the mean of 500 bootstrap samples of $R$ and its 95\% confidence interval, respectively. However, it can be seen that the performance of the bootstrap confidence intervals are quite well in both cases, but construction of these intervals are computationally more expensive than the asymptotic confidence intervals.
\par
Based on the Bayesian method described in \ref{Bayes2}, the Bayes estimate of the stress-strength parameter when the hyper-parameters of the prior distributions are set at $\gamma_{i}=0.1, \lambda_{i}=5,~i=1,\ldots,4$ is $\hat{R}_{B}=0.6282$, and the associated
95\% credible interval is (0.4974,  0.7529). This Bayes estimate is quite robust with respect to the changes in values of the hyper-parameters, even if we select a non-informative prior distribution by setting $\gamma_{i}=0.0001, \lambda_{i}=0.0001,~i=1,\ldots,4$, at which the range of possible changes of $\hat{R}_{B}$ is adequately small. This Bayes estimate could be sensitive to the changes of the prior distributions, if a very small value chosen for the shape parameter(s) of the Gamma distributions against a quite large value chosen for the scale parameter(s) of the Gamma distributions introduced above.
\par
Table \ref{tab5} illustrates the MLEs and Bayes estimates based on the simulated data associated with the given parameters. We use the methods presented in Section \ref{general} to calculate these estimations.
\par
We now perform some numerical experiments based on the censored samples under different progressive censoring schemes. For simplicity, we assume the scale parameters are common and known in both strength and stress distributions. Therefore, to simulate the data, we take $a=1, b=1.5$ and $\alpha=1.5$ and $\beta=0.5$. For a given $n$ and $m$, three different progressive censoring schemes are used to generate the
data: (i) the usual type-II censoring scheme, where $n-m$ remaining items are removed at the $m$-th failure; (ii) type-III censoring
scheme (in this scheme, $n-m$ items are randomly discarded at the first failure); (iii) type-IV censoring scheme, at which all the $P_i$'s are taken the same number. Note that the first two censoring schemes are extreme ones, but the last censoring scheme lies in between the first two.
\par
For given $(n_{1},m_{1})$ and progressive censoring scheme $\{n_{1},m_{1},P_{1},\ldots,P_{m_{1}}\}$ for the first variable, and $(n_{2}, m_{2})$ and progressive censoring scheme $\{n_{2},m_{2},Q_{1},\ldots,Q_{m_{2}}\}$ for the second variable, the simulation is replicated 1000 times. In each simulation, the MLE of $R$, the average of biases and the mean squared errors (MSEs) for this estimator are calculated based on the simulated data and reported in Table \ref{tab6}.
\par
Based on the results reported in Table \ref{tab6}, it is clear that MLE performs quite well. In addition, it is observed that as the sample sizes increase the performances become better.

\begin{table}
\begin{center}
\begin{tabular}{l l l l l l }
 \hline
 (n,m) & $R$ & $\hat{R}$ & $\hat{R}_{B}$ &  $CI_{MLE}$ & $cp$ \\ \hline
   (15,15)&0.416&0.421&0.412&(0.261,0.581)&0.952 \\
    &   0.500 & 0.498 &0.5025 &(0.282,0.721)&0.956\\
    &   0.583 &0.583 &0.5833 &(0.459,0.715)&0.923\\
    &   0.666 &0.662 & 0.6664 &(0.340,0.976)&0.788\\ \hline
    (25,25)&0.416&0.418&0.415&(0.301,0.535)&0.979 \\
    &   0.500 & 0.504 &0.4976 &(0.379,0.629)&0.977\\
    &   0.583 & 0.583 &0.5829&(0.443,0.723)&0.972\\
    &   0.666 & 0.662 & 0.6658 &(0.459,0.866)&0.931\\ \hline
(25,25)&0.416&0.419&0.4158&(0.271,0.567)&0.976 \\
    &   0.500 & 0.498 &0.4992 &(0.353,0.642)&0.983\\
    &   0.583 & 0.583 &0.583&(0.452,0.713)&0.989\\
    &   0.666 & 0.668 & 0.6659 &(0.526,0.809)&0.984\\ \hline
\end{tabular}
\caption{Simulation results and estimation of the parameters when
$a,b$ are known from 1000 samples.}\label{Tab1}
\end{center}
\end{table}

\begin{table}
\begin{center}
\begin{tabular}{l l l l l l l}
 \hline
 (n,m) & $(a,b,\alpha,\beta)$ & $(\hat{a},\hat{b},\hat{\alpha},\hat{\beta})$ & $Bias(R)$ &  $MSE(\hat{R})$ & $CI$ & $cp$\\

\hline (15,15)&(0.5,0.5,1,1)&(0.5407,0.5578,1.1371,1.1229)&0.0019&0.0093&(0.595,1.4049)&0.793 \\
    &   (1.5,0.5,2,1.5) & (1.6503,0.6001,2.5363,1.8865) &-0.0036 &0.0092&(0.2976,0.8509)&0.632\\
    &   (2,1.5,0.5,0.5) & (2.1562,2.3411,0.5467,0.5567) &-0.0021 & 0.0086& (0.3189  0.6842)&0.951\\\vspace{8pt}
    &   (2,1.5,3,2.5) & (2.2721,1.5117,4.3499,3.5099) & -0.0018 & 0.009&(0.0359,1.1268)&0.711\\

    (25,25) &   (0.5,0.5,1,1)&(0.5608,0.5114,1.1075,1.1201)&-0.0023&0.0052&(0.0552,1.6174)&0.909\\
    &   (1.5,0.5,2,1.5) & (1.5693,0.5812,2.2741,1.7148) & -0.0055&0.0052&(0.2468,0.8979)&0.847\\
    &   (2,1.5,0.5,0.5) & (2.0581,1.9989,0.5292,0.5308)&0.0002&0.0054&(0.3574,0.6404)&0.95\\\vspace{8pt}
    &   (2,1.5,3,2.5) & (2.1289,1.5520,3.6893,2.9854) &0.004&0.0049&(0.0102,1.1067)&0.773\\

(25,50) &(0.5,0.5,1,1)&(0.5373,0.5005,1.0856,1.0640)&0.0016&0.0040&(0.3198,0.6845)&0.951\\
    &   (1.5,0.5,2,1.5) & (1.4969,0.5983,2.2089,1.5589) &0.0075&0.0039&(0.043,1.1045)&0.98\\
    &   (2,1.5,0.5,0.5) & (2.0272,1.8624,0.5272,0.5090)&0.0054&0.004&(0.3808 ,0.6247)&0.942\\\vspace{8pt}
    &   (2,1.5,3,2.5) & (1.9882,1.6503,3.3075,2.7007) &0.0001&0.0038&(0.633,0.882 )&0.815\\

(50,50)
&(0.5,0.5,1,1)&(0.5409,0.4928,1.0543,1.0739)&-0.0045&0.0026&(0.2342 ,0.7684)&0.959\\
    &   (1.5,0.5,2,1.5) & (1.4993,0.5744,2.1120,1.5750) &-0.0007&0.0025&(0.3748,0.7698)&0.997\\
    &   (2,1.5,0.5,0.5) & (2.0291,1.7810,0.5109,0.5121)&-0.0004&0.0026 &(0.4022,0.6005)&0.972\\
    &   (2,1.5,3,2.5) & (2.0067,1.6199,3.2356,2.6983) &-0.0013 &0.0026 &(0.2194,0.8787 )&0.852\\

\hline
\\
\end{tabular}
\caption{Simulation results and estimation of the parameters when
$a,b$ are unknown from 1000 samples.}\label{tab2}
\end{center}
\end{table}


\begin{table}
\begin{center}
\begin{tabular}{l l l l l|| l l l l l}
 \hline
  &  & $x$ &  & & & & $y$ &  & \\ \hline
  1.5671 &  1.9729 & 0.2760 & 2.0204 & 1.0948 &  1.3314 &  1.7499 &  0.1323 & 1.7989 &   0.8546\\
   0.2278 & 0.5049 &  0.9314 &    2.4024& 2.4994 & 0.1006 &  0.3075 & 0.6950 &  2.1949 & 2.2954 \\
    0.3240 &  2.5877 & 2.3983 &  0.8252 &   1.5189 & 0.1660 &  2.3870 &  2.1907 &   0.5938 &  1.2821\\
    0.2996 &  0.7221 &  2.0358 &  1.4932 &  2.4269 &  0.1486 &  0.4981 &  1.8150 &  1.2558 &   2.2203\\ \hline
\end{tabular}
\caption{The data generated with $n = m = 20$, $a = 1,b = 0.4 ,\alpha=1.5$ and $\beta=1$.}\label{tab3}
\end{center}
\end{table}
\begin{table}
\begin{center}
\begin{tabular}{l l l l l}
 \hline
 $a,b$ & $(\hat{a},\hat{b},\hat{\alpha},\hat{\beta})$ & $\hat{R}$ & $\hat{R}^{\ast}$  & $CI_{boot}$ \\ \hline
    Unknown &(1.12,0.47,1.12,0.61) &0.6449 & 0.6468&(0.4777,0.7976)\\
    Known   &(1,0.4,1.64,0.84) &0.7283 & 0.7275&(0.6064,0.8442)\\
\hline
\end{tabular}
\caption{Parameters estimation and bootstrap confidence intervals
with $N =500$ boot times for the data presented in Table \ref{tab3}.}\label{tab4}
\end{center}
\end{table}

\begin{table}
\begin{center}
\begin{tabular}{l l l l l l l l l l}
 \hline (n,m) & $a_{1}$ & $b_{1}$ & $a_{2}$ & $b_{2}$ & $\alpha$ & $\beta$ &$R$ & $\hat{R}$  &$\hat{R}_{B}$\\

\hline (25,25)&1&0.5&1.5&0.5&1.5&1&0.6922&0.7010 &0.6994 \\
    &1&0.5&1.5&0.5&1.5&1.5&0.6071&0.6323& 0.6327\\
    &0.7&0.5&1&0.15&1.5&1.5&0.5315&0.4905 &0.5111\\\vspace{8pt}
     &1.5&1.5&1.5&1.5&1&2&0.3333&0.3558&0.3540\\

    (25,50)&1&0.5&1.5&0.5&1.5&1&0.6922&0.7043&0.6967\\
    &1&0.5&1.5&0.5&1.5&1.5&0.6071&0.6325&0.6212\\
    &0.7&0.5&1&0.15&1.5&1.5&0.5315&0.5076&0.5224\\\vspace{8pt}
     &1.5&1.5&1.5&1.5&1&2&0.3333&0.3442&0.3437\\

(50,50)&1&0.5&1.5&0.5&1.5&1&0.6922&0.6883&0.6933\\
    &1&0.5&1.5&0.5&1.5&1.5&0.6071&0.6189&0.6154\\
    &0.7&0.5&1&0.15&1.5&1.5&0.5315&0.5350&0.5287\\\vspace{8pt}
     &1.5&1.5&1.5&1.5&1&2&0.3333&0.3271&0.3358\\ \hline
\end{tabular}
\caption{Simulation results and estimation of $R$ in general
case.}\label{tab5}
\end{center}
\end{table}

\begin{table}
\begin{center}
\begin{tabular}{l l l l l l l l}
 \hline
 $(n_{1},m_{1})$ & $(n_{2},m_{2})$ & $P$ & $Q$ &  $Bais(\hat{R})$ &  $MSE(\hat{R})$  \\

\hline (10,5)&(10,5)&II&II&-0.0116&0.0169  \\
       (10,5)&(10,5)&II&III&-0.019&0.0159\\
       (10,5)&(10,5)&III&II&-0.0236&0.0176 \\
       (10,5)&(10,5)&III &III&-0.0209&0.018 \\\vspace{8pt}
       (10,5)&(10,5)&IV &IV&-0.0187&0.0164 \\

       (20,5)&(20,5)&II&II&-0.0204&0.0222  \\
       (20,5)&(20,5)&II&III&-0.0307&0.0217 \\
       (20,5)&(20,5)&III&II&-0.0278&0.021   \\
       (20,5)&(20,5)&III &III&-0.0253&0.0208  \\\vspace{8pt}
       (20,5)&(20,5)&IV &IV&-0.0127&0.0159   \\

       (20,10)&(20,10)&II&II&-0.0162&0.0104  \\
       (20,10)&(20,10)&II&III&-0.0123&0.0091 \\
       (20,10)&(20,10)&III&II&-0.006&0.0091 \\
       (20,10)&(20,10)&III &III&-0.0101&0.0094 \\\vspace{8pt}
       (20,10)&(20,10)&IV &IV&-0.0097&0.0081 \\

       (30,10)&(30,10)&II&II&-0.0131&0.0128   \\
       (30,10)&(30,10)&II&III&-0.0176&0.0126  \\
       (30,10)&(30,10)&III&II&-0.0121&0.011 \\
       (30,10)&(30,10)&III &III&-0.0149&0.0129  \\\vspace{8pt}
       (30,10)&(30,10)&IV &IV&-0.0097&0.0077  \\

       (30,15)&(30,15)&II&II&-0.0035&0.0061  \\
       (30,15)&(30,15)&II&III&-0.0078&0.0071 \\
       (30,15)&(30,15)&III&II&-0.008&0.0061  \\
       (30,15)&(30,15)&III &III&-0.0088&0.0068  \\\vspace{8pt}
       (30,15)&(30,15)&IV &IV&-0.0074&0.005 \\
\hline
\end{tabular}
\caption{Biases and MSEs for MLEs and Bayes estimates under different censoring schemes when
$a=1$, $b=1.5$, $\alpha=1.5$ and $\beta=0.5$.}\label{tab6}
\end{center}
\end{table}

\section{Data Analysis}
In this section, we apply the procedures presented in this paper to estimate $R$ on a real life case study. The data shown in Tables \ref{tab7} and \ref{tab8} are the breaking strengths of jute fiber at two different gauge lengths. These data were first used by Xia et al. (2009) and were then re-used by Sara\c{c}o\v{g}lu et al (2012) to study the estimation of the stress-strength parameter for Exponential distribution under progressive type-II censoring.
\par
\begin{table}
\begin{center}
\begin{tabular}{l l l l l l l l}
 \hline
    693.73 & 704.66 & 323.83 & 778.17 &123.06 &637.66 &383.43& 151.48\\
108.94 & 50.16 & 671.49 & 183.16 &257.44 &727.23& 291.27& 101.15\\
376.42& 163.40& 141.38& 700.74& 262.90& 353.24& 422.11& 43.93\\
590.48& 212.13& 303.90& 506.60& 530.55& 177.25 & &\\ \hline
\end{tabular}
\caption{Breaking strength of jute fiber of gauge length 10 mm (here denoted by $X$ variable).}\label{tab7}
\end{center}
\end{table}

\begin{table}
\begin{center}
\begin{tabular}{l l l l l l l l}
 \hline
    71.46& 419.02& 284.64& 585.57& 456.60& 113.85& 187.85& 688.16\\
662.66& 45.58& 578.62& 756.70& 594.29& 166.49& 99.72& 707.36\\
765.14& 187.13& 145.96& 350.70& 547.44& 116.99& 375.81& 581.60\\
119.86& 48.01& 200.16& 36.75& 244.53& 83.55& & \\ \hline
\end{tabular}
\caption{Breaking strength of jute fiber of gauge length 20 mm ((here denoted by $Y$ variable).}\label{tab8}
\end{center}
\end{table}
We also use these data to estimate the stress-strength model when the following distributions GLFRD($a,b,\alpha$) and GLFRD($a,b,\beta$) are fitted to the data given in Tables \ref{tab7} and \ref{tab8}, respectively. The maximum likelihood estimators of $a,b,\alpha$ and $\beta$ are 0.0027, $2.4352\times10^{-6}$, 1.6185 and 1.3209, respectively. The Kolmogorov-Smirnov statistics' values are 0.093 and 0.1297 and the corresponding $p$-values are 0.9363 and 0.6468, respectively. The derived $p$-values indicate that GLFR distributions with the estimated parameters are fitted very well to the data. Based on the complete data, we then compute the MLE of $R = P(Y < X)$  which is $\hat{R}= 0.5506$ and the associated 95\% confidence interval is (0.4232, 0.6717). Using the methods explained in Subsection 4.4, the Bayes estimate of $R$ with respect to improper priors is $\hat{R}_{B}= 0.5517$, and the associated 95\% credible interval is $(0.4418, 0.6544)$.
\par
We then consider three different progressively censored samples which have been generated from the above data sets with $m_{1} = m_{2} = 15$ in all the cases: (i) Scheme-1: (type-II, type-II), (ii) Scheme-2: (type-III, type-III) and (iii) Scheme-3 (type-IV, type-IV). The MLEs, their corresponding 95\% confidence intervals are reported in Table \ref{tab9}. Clearly,  the estimated $R$ obtained using the third scheme is closer to the estimates obtained based on the complete data.
\par
\begin{table}[h]
\begin{center}
\begin{tabular}{l l l l }
 \hline
 Scheme & $\hat{R}$ & 95\% CI  \\ \hline
Scheme-1 &0.5791 & (0.3993, 0.7401) \\
Scheme-2  &0.6445 &  (0.4669, 0.7896)\\
Scheme-3  &0.5298 & (0.3525, 0.6999) \\ \hline
\end{tabular}
\caption{The MLEs, the associated 95\% confidence intervals of $R$ under the different censoring schemes.}\label{tab9}
\end{center}
\end{table}

\section{Conclusions}
In this paper, we have investigated the issue of estimating the stress-strength parameter for the generalized linear failure rate distribution in the different situations when the scale parameters of the stress and strength distributions are either common and known or unknown, and also in the general case in which the parameters of these two distributions are not common and known. The potential flexibility of the GLFR distribution is the main reason behind the study carried out in this paper. The generalized linear failure rate distribution introduced by Sarhan and Kundu (2007), developed by Sarhan et al (2008) and recently studied in more depth by Shahsanei (2011) is the generalisations of the linear failure rate distribution, generalized exponential and generalized Rayleigh, and many more distributions. In addition, by using this distribution, the lifetime data with the different patterns for the hazard rate functions, including increasing, decreasing and bathtub shaped can be statistically studied.
\par
When the common scale parameters are common and known, the Bayes estimate slightly performs than the MLE, in the sense that its bias and MSE is smaller. The MLE of $R=P(Y<X)$ is quite straightforward, and two approximate Bayes estimators based on 0-1 and squared error loss functions are also presented at which they show similar performance. Similar to the previous relevant studies, we can also derive the asymptotic distribution of the MLE to construct the associated confidence intervals which also work quite well.
\par
The computation of the MLE of $R$ when the common scale parameters are unknown can be done using an iterative numerical introduced above. The corresponding asymptotic distribution of the MLE of $R$ is then obtained using the Delta method given in the Appendix. It is then trivial to construct an asymptotic confidence interval based on this distribution. We also derive the parametric percentile Bootstrap confidence interval for any sample sizes, in particular, for small sample sizes. We show that this interval illustrates satisfactory performance in practice even for very small sample sizes. As expected, to compute the Bayes estimators in this case, one must use the expensive simulation methods described above. One of these simulation methods which is more common here and is based on the acceptance rejection sampling procedure and originally proposed by Devroye (1984). This method enables us to generate a sample from the posterior distribution of $R$. As a result, the Bayes estimate of $R$ along with the associated credible interval can be easily computed using the simulated data. An alternative way is to use the Empirical Bayes method where we first substitute the estimates of the scale parameters (the MAP estimations used here) into the posterior distribution of $R$, and we then can calculate the Bayes estimates of $R$ under the 0-1 or squared error loss functions, as mentioned in 3.2.
\par
We have used the similar methods to calculate the MLE, the related confidence intervals, the Bayes estimates and its associated credible intervals for the general case where the parameters of the both variables are neither common and nor known.
\par
Finally, we have addressed the estimation problem of $R=P(Y<X)$ when the observed data are progressively type-II censored samples from both GLFR distributions. We have only reported the MLEs of $R$ and the associated confidence intervals when the common scale parameters are known under three censoring schemes described above. However, it is trivial to extend this study to obtain the MLEs of $R$ for more general cases, but the Bayes estimates of $R$ in these situations required more sophisticated MCMC based simulation methods. This work is in progress, and it will be reported later.
\par
We are also developing this study by applying recent advances in Bayesian inference based on higher-order asymptotic, pseudo-likelihoods, and related matching priors, which allow one to perform more accurate inference on the stress-strength parameter, even for small sample sizes. In addition, these approaches have the advantages of avoiding the elicitation on the nuisance parameters and the expensive computation of multidimensional integrals.

\section{Appendix}

\textbf{Proof of Theorem 2.}~~We use the Delta method to prove this theorem. In the Delta method, it is stated that if $g:\mathbb{R}^{k}\to \mathbb{R}^{l}$ has a  derivative $ \nabla g(\textbf{a})$ at $\textbf{a}\in \mathbb{R}^{k}$ and
\[
n^{b}\{\textbf{X}_{n}-\textbf{a}\}\to^{\!\!\!\!\!\!{d}}\textbf{Y}
\]
for some $k$-vector $\textbf{Y}$ and some sequence $\textbf{X}_{1},\textbf{X}_{2},\ldots$ of $k$-vectors, where $b>0$, then
\[
n^{b}\{g(\textbf{X}_{n})-g(\textbf{a})\}\to^{\!\!\!\!\!\!{d}}~~[\nabla g(\textbf{a})]^{T}\textbf{Y}.
\]
By Theorem 1, we know that as $n\to \infty$, $m\to \infty$ and $\frac{n}{m}\to p$, then
\[
\left(
  \begin{array}{c}
    \sqrt{n}(\hat{a}-a) \\
    \sqrt{n}(\hat{b}-b) \\
    \sqrt{n}(\hat{\alpha}-\alpha) \\
    \sqrt{m}(\sqrt{\frac{m}{n}}\hat{\beta}-\sqrt{\frac{m}{n}}\beta)\\
  \end{array}
\right)
\rightarrow N_{4}({\bf 0},{\bf U^{-1}}(a,b,\alpha,\beta))
\]
where $\textbf{U}$ is given in (\ref{U-equ}).
\par
The $\textbf{U}^{-1}$ denote the covariance matrix of the Multivariate normal distribution given above and can be factorised as $\textbf{U}^{-1}=k^{-1}A$, where $A$ is the adjoint matrix of $U$, $k$ is the determinant of $U$, and are both given, respectively, by
\[
A=\left(
    \begin{array}{cccc}
      a_{11} & a_{12} & a_{13} & a_{14} \\
       & a_{22}& a_{23} & a_{24} \\
       &  & a_{33} & a_{34} \\
       &  &  & a_{44} \\
    \end{array}
  \right),
\]
where
\[
a_{11}=u_{22}u_{33}u_{44}+u_{23}u_{34}u_{42}+u_{24}u_{32}u_{43}-u_{22}u_{34}u_{43}-u_{23}u_{32}u_{44}-u_{24}u_{33}u_{42}
\]
\[
a_{12}=u_{13}u_{32}u_{44}+u_{14}u_{33}u_{42}-u_{13}u_{34}u_{42}-u_{14}u_{32}u_{43}
\]
\[
a_{13}=u_{13}u_{24}u_{42}+u_{14}u_{22}u_{43}-u_{13}u_{22}u_{44}-u_{14}u_{23}u_{42}
\]
\[
a_{14}=u_{13}u_{22}u_{34}+u_{14}u_{23}u_{32}-u_{13}u_{24}u_{32}-u_{14}u_{22}u_{33}
\]
\[
a_{22}=u_{11}u_{33}u_{44}+u_{13}u_{34}u_{41}+u_{14}u_{31}u_{43}-u_{11}u_{34}u_{43}-u_{13}u_{31}u_{44}-u_{14}u_{33}u_{41}
\]
\[
a_{23}=u_{11}u_{24}u_{43}+u_{14}u_{23}u_{41}-u_{11}u_{23}u_{44}-u_{13}u_{24}u_{41}
\]
\[
a_{24}=u_{11}u_{23}u_{34}+u_{13}u_{24}u_{31}+u_{14}u_{21}u_{33}-u_{11}u_{24}u_{33}-u_{14}u_{23}u_{31}
\]
\[
a_{33}=u_{11}u_{22}u_{44}-u_{11}u_{24}u_{42}-u_{14}u_{22}u_{41}
\]
\[
a_{34}=a_{34}=u_{11}u_{24}u_{32}+u_{14}u_{22}u_{31}-u_{12}u_{24}u_{31}-u_{14}u_{21}u_{32}
\]
\[
a_{44}=u_{11}u_{22}u_{33}+u_{12}u_{23}u_{31}+u_{13}u_{21}u_{32}-u_{11}u_{23}u_{32}-u_{12}u_{21}u_{33}-u_{13}u_{22}u_{31}
\]
and \[
k=u_{11}u_{22}u_{33}u_{44}+u_{11}u_{23}u_{34}u_{42}+u_{11}u_{24}u_{32}u_{43}+u_{13}u_{22}u_{34}u_{41}+u_{13}u_{24}u_{31}u_{42}+
\]
\[
u_{14}u_{22}u_{31}u_{43}+u_{14}u_{23}u_{32}u_{41}-u_{11}u_{22}u_{34}u_{43}-u_{11}u_{23}u_{32}u_{44}-u_{11}u_{24}u_{33}u_{42}-
\]
\[
u_{13}u_{22}u_{31}u_{44}-u_{13}u_{24}u_{32}u_{41}-u_{14}u_{22}u_{33}u_{41}-u_{14}u_{23}u_{31}u_{42}.
\]
Now, we define a function associated with $R$ as, $R:(\mathbb{R}^{+})^{4}\to (0,1)$, such that $R$ applied to the vector of $(a,b,\alpha,\beta)$, yields $R=\alpha/(\alpha+\sqrt(\frac{n}{m})B_{1})$ (where, $B_{1}=\frac{m}{n}\beta$). Then, using the Delta Theorem
\[
\nabla R\left(
        \begin{array}{c}
          a \\
          b \\
          \alpha \\
          B_{1} \\
        \end{array}
      \right)=\left(
                \begin{array}{c}
                  0 \\
                  0 \\
                  \sqrt(\frac{n}{m})B_{1}/[(\alpha+\sqrt(\frac{n}{m})B_{1})]^{2} \\
                   -\sqrt(\frac{n}{m})\alpha/[(\alpha+\sqrt(\frac{n}{m})B_{1})]^{2} \\
                \end{array}
              \right)
\]
Therefore, the variance given in (\ref{Theorem2-sigma}) will be achieved as followed
\[
{\sigma^{2}}^{*}=\left[\nabla R\left(
        \begin{array}{c}
          a \\
          b \\
          \alpha \\
          B_{1} \\
        \end{array}
      \right)\right]^{T} U^{-1}(a,b,\alpha,\beta)\left[\nabla R\left(
        \begin{array}{c}
          a \\
          b \\
          \alpha \\
          B_{1} \\
        \end{array}
      \right)\right]=
      \]
      \[
      \frac{1}{k(\alpha+\beta)^{4}}[\beta^{2}a_{33}-2\sqrt{p}\alpha\beta a_{34}+\alpha^{2}{p}a_{44}]
\]

\end{document}